\documentclass[preprint,12pt,authoryear]{elsarticle}

\usepackage{amssymb}
\usepackage{amsmath}

\usepackage{amsfonts}
\usepackage{graphicx}
\usepackage{bbm}
\usepackage{bbold}
\usepackage{mathrsfs}
\usepackage{calligra}
\usepackage{multirow}
\usepackage{subfig}
\usepackage{enumitem}
\usepackage{setspace}
\usepackage[OT4]{fontenc}
\usepackage{xcolor}
\newtheorem{theorem}{Theorem}
\newtheorem{proposition}{Proposition}
\newtheorem{remark}{Remark}

\begin{document}

\begin{frontmatter}



\title{Detecting dependence structure: visualization and inference} 


\author[BC]{Bogdan \'Cmiel} 
\ead{cmielbog@gmail.com}
\author[TL]{Teresa Ledwina}
\ead{ledwina@impan.pl}
\address[BC]{AGH University of Krakow, Faculty of Applied Mathematics,\\ Al. Mickiewicza 30, 30-059 Krak\'ow, Poland}
\address[TL]{Institute of Mathematics, Polish Academy of Sciences,\\ ul. Kopernika 18, 51-617 Wroc{\l}aw, Poland}

\begin{abstract}
Identifying dependency between two random variables is a fundamental problem. The clear interpretability and ability of a procedure to provide information on the form of possible dependence is particularly important when exploring dependencies. In this paper, we introduce a novel method that employs a new estimator of the quantile dependence function and pertinent local acceptance regions. This leads to an insightful visualisation and a rigorous evaluation of the underlying dependence structure. We also propose a test of independence of two random variables, pertinent to this new estimator. Our procedures are based on ranks, and we derive a finite-sample theory that guarantees the inferential validity of our solutions at any given sample size. The procedures are simple to implement and computationally efficient. The large sample consistency of the proposed test is also proved. We show that, in terms of power, the new test is one of the best statistics for independence testing when considering a wide range of alternative models. Finally, we demonstrate the use of our approach to visualise dependence structure and to detect local departures from independence through analysing some real-world datasets.
\end{abstract}


\end{frontmatter}


\newpage
\topmargin=-2cm
\textwidth 6.7in
\textheight 24cm
\hoffset -1.7cm
\noindent
{\bf Content}\\

\noindent
{\it Paper}\\

\noindent
1. Introduction \dotfill {3} \\
2. Quantile dependence function $q$ \dotfill {4} \\
3. Empirical checkerboard copula, new estimator of the measure $q$, and their properties \dotfill {6} \\
\hspace*{0.5cm} 3.1. Bilinear estimator $\bar{C}_n$ of $C$ \dotfill {7} \\
\hspace*{0.5cm} 3.2. New estimator $\bar{q}_n$ of $q$ \dotfill {8} \\
\hspace*{0.5cm} 3.3. Illustration \dotfill {8} \\
4. Inspection of dependence structure and first illustrations \dotfill {10} \\
\hspace*{0.5cm} 4.1. Discretization, simultaneous acceptance regions, and the dependence diagram \dotfill {11} \\
\hspace*{0.5cm} 4.2. Preliminary comparison of the approach of Zhang's (2019) to ours \dotfill {13} \\
\hspace*{0.5cm} 4.3. Graphical analysis of two popular data sets \dotfill {13} \\
5. New test of global independence \dotfill {16} \\
\hspace*{0.5cm} 5.1. Test statistic and its properties \dotfill {16} \\
\hspace*{0.5cm} 5.2. Power study - illustrative cases \dotfill {17} \\
\hspace*{0.5cm} 5.3. Application to COVID-19 infection dynamics data \dotfill {20} \\
6. Multivariate variants of $q$ and $\bar{q}_n$ \dotfill {22} \\
7. Conclusion and discussion \dotfill {23} \\
References \dotfill {24} \\
\\
\\
{\it Supplementary Material}\\

\noindent
A.1. Full list of alternatives, empirical powers, and related comments \dotfill\ 1 \\
A.2. Some comments on the sensitivity of empirical powers to the grid size $d(n)$ and the smoothing parameter $t(n)$ \dotfill\ 3 \\
A.3. The values of barriers in selected real data examples and their stability \dotfill\ 8 \\
A.4. Proofs \dotfill\ 10 \\
\hspace*{0.5cm} A.4.1. Proof of Proposition 1 \dotfill\ 10 \\
\hspace*{0.5cm} A.4.2. Proof of Proposition 2 \dotfill\ 10 \\
\hspace*{0.5cm} A.4.3. Proof of Theorem 1 \dotfill\ 11 \\
A.5. Zhang's BET framework. New interpretation and related comments \dotfill\ 12 \\
A.6. Brief comments on some approaches to measuring association \dotfill\ 14 \\
References for Appendices \dotfill\ 15

\newpage
\section{Introduction}\label{s1}
The problem of testing independence and measuring the strength of dependence is one of the
basic inference tasks and has been studied for more than a century. Early works have focused
on proposing some measures and related test statistics, and discussing possibilities of their use.
Later, after the development of probabilistic and computational tools, considerable attention
was focused on the comparison of existing procedures and understanding their abilities and
weaknesses. The next step was some effort to extend the range of sensitivity of the classical tests.
In recent years, new tasks have been formulated. They include quantification and estimation
of dependency, learning the dependency structure and its visualisation, and detecting local
discrepancies or some particular structures, among others. In the following, we briefly summarise some recent developments in the area.

In the past twenty-five years or so, there has been a significant surge of ideas for constructing independence tests. One of the stimulating
factors could be interest in the problem not only among statisticians but also among engineers
and data science community, as well. In addition, the need to analyse
some massive sets of data and multivariate observations has stimulated considerable progress.
These factors have resulted in a variety of approaches to
the problem. We mention some directions: data driven tests (\cite{Kallenberg}), tests based on the empirical copula process (\cite{Genest04}), approaches based on partitions (\cite{Thas}, \cite{Reshef11,Reshef16}, \cite{Heller13, Heller16}), locally optimal rank tests (\cite{Genest05}), kernel-based methods (\cite{Gretton}, \cite{Schrab}), 
seminorms of some weighted empirical processes (\cite{Deheuvels}, \cite{Berghaus}, \cite{Cmiel}), 
Brownian distance covariance (\cite{Szekely}), local
Gaussian correlation (\cite{Berentsen}), sign covariance (\cite{Bergsma}), empirical likelihood approach for densities (\cite{Vexler14}), mutual information
(\cite{Suzuki}, \cite{Berrett}, \cite{Gonzalez}), Bernstein empirical
copula based solutions (\cite{Belalia}), generalized R-squared (\cite{Wang}),
point processes perspective (\cite{Dvorak}), matching bipartite ranking and ROC analysis (\cite{Limnios24}).
The next ideas are discussed below.

Beyond the task of testing global independence, some authors have investigated the interpretation
and quantification of dependency in the case of rejection of independence. The
question naturally relates to some local dependence measures, which, however, are mostly used
as explanatory tools, similarly as chi-plots (\cite{Fisher85}) and Kendall’s plots (\cite{Genest03}), which are standard in the field. Presumably, the first local measure discussed and applied to identify departures from independence
was that of \cite{Holland}. \cite{Jones}, starting from an estimate of
this measure, introduced the related dependence map with domain in ${\mathbb R}^2$. A similar approach
has been elaborated in \cite{Berentsen}, in application to another measure,
the local Gaussian correlation. Both papers introduce some grids of points in ${\mathbb R}^2$ and search
for global regions where pertinent estimated measures are too big (too small). These regions
define related dependence maps. Other solutions were proposed by \cite{Ma}, \cite{Zhang}, and \cite{Lee}. These three papers exploit some two-dimensional grids in $[0, 1]^2$,
deal with diadic partitions, and refer to some testing procedures pertinent to likelihood in
adequately defined contingency tables. In the case of rejection of independence, these works
provide some likely patterns of dependency. 
 
In our paper, we address the two issues discussed above. 

The first of our contribution is a set of graphical tools that facilitate a comprehensive view of the association structure and the magnitude of the underlying dependence. To be specific, we consider a bivariate random vector with
continuous marginal cumulative distribution functions and our starting point is a properly standardised
copula, i.e. the quantile dependence function $q$, defined on $(0, 1)^2$, as introduced
in \cite{Ledwina14arx,Ledwina15}. This is a function-valued measure of dependence equal to 0 in the
case of independence. We propose and discuss a simple and useful rank estimator $\bar q_n$ of
$q$. This estimator inherits all properties after $q$ and provides a detailed and easily interpretable
visualisation of the deviations from independence. To formally assess the significance of
the magnitude of observed discrepancies in the plot, we introduce a grid of bivariate diadic
partitions of $(0, 1)^2$ and related local simultaneous acceptance regions. They define our variant
of the dependence map, which we call the {\it dependence diagram}, by showing the regions of
pairs of empirical quantiles where the differences between observed  $\bar q_n$   and expected under independence are statistically significant. The results of this approach nicely confront the evidence provided by Zhang's BET (\cite{Zhang}).

The second contribution of this paper is a new independence
test statistic, also clearly related to $\bar q_n$, which estimates the global maximum of $\bar q_n$ on the grid. This appears to be a competitive test for the global independence hypothesis, which, over a wide range of alternative models,
has a high and stable power similar to that of the best of existing solutions. 

Our approach is simple, fully
nonparametric, and requires only standard Monte Carlo simulations under independence. In
addition, it is not only powerful when dealing with testing for independence, but provides simultaneously
insightful evidence and interpretable description of departures from the null model.
The proposed tools are flexible and capable of revealing and capturing complex dependence
structures.

Basic facts about the quantile dependence function are presented in the next section. The
new estimator of the function is given in Section \ref{s3}. Local acceptance regions are discussed in
Section \ref{s4}. The proposed procedure is illustrated with some real data examples; see Sections \ref{s4} and \ref{s5}. The new global test construction is given in Section \ref{s5} and is accompanied by a small
representative simulation study. The multivariate extensions of $q$ and $\bar{q}_n$ are presented in Section \ref{s6}. Section \ref{s7} concludes the paper. Furthermore, Supplementary Material, consisting of six Appendices, provides extensive comparison of the performance of our proposed test and several competitors on various synthetic data sets, practical implementation questions, all proofs, as well as further comments on recent Zhang's diagnostic tool. We conclude with brief  remarks on some earlier approaches to measuring the degree of dependence.

\section{Quantile dependence function $q$}

Consider a pair of random variables $X$ and $Y$ with a bivariate cumulative distribution function (cdf) $H$ and with continuous margins $F$ and $G$, respectively. Denote this class of bivariate cdf's by ${\cal H}^c$. For elements of ${\cal H}^c$ there exists a unique copula $C$ such that $C(u,v)=H(x_u,y_u)$, where $x_u=F^{-1}(u),\ y_v=G^{-1}(v),\ (u,v) \in [0,1]^2$. Moreover, let the function $\mathcal{I}(u,u)=uv$ denote the independence copula function.

\cite{Ledwina14arx,Ledwina15} has proposed measuring the discrepancy between the copula $C$ and the independence copula $\mathcal{I}$ by 
\begin{equation}\label{1}
q(u,v)=\frac{C(u,v)-uv}{\sqrt{uv(1-u)(1-v)}}=\frac{H(x_u,y_v)-F(x_u) G(y_v)}{\sqrt{uv(1-u)(1-v)}},\;\;\;(u,v) \in (0,1)^2.
\end{equation}
For successive marginal quantiles $(u,v)$, the measure $q$ quantifies an allocation of the probability mass under cdf $C$ in comparison to that pertinent to the independence copula $\mathcal{I}$. Given $(u,v)$, the normalisation makes natural estimators of $q$ asymptotically $N(0,1)$ when the true underlying copula is the independence copula $\mathcal{I}$. The factor $[uv(1-u)(1-v)]^{-1}$ is the asymptotic variance of the classical empirical copula built from observations obeying cdf $\mathcal{I}$; see also Section \ref{s3.1}. 

The definition of $q$ comes from earlier attempts to investigate the structure of $C$ by looking at successive correlation coefficients of some rich enough class of functions of the transformed variables $F(X)$ and $G(Y)$; cf. \cite{Kallenberg},  \cite{Janic}, and \cite{Ledwina14}. In case of (\ref{1}), we  see that for $(u,v) \in (0,1)^2$ it holds that
\begin{equation}\label{2}
q(u,v)=\text{Corr}[h_u(F(X)),h_v(G(Y))],\;\;\mbox{where}\;\;h_u(w)=\frac{u-\mathbb{1}(0 \leq w \leq u)}{\sqrt{u(1-u)}},\;\;w \in (0,1),
\end{equation}
and $h_v(\cdot)$ is defined analogously, where $\mathbb{1}(A)$ denotes the indicator of the set $A$. \cite{Ledwina12} have derived the system $\{h_v(\cdot)\}$ starting from projections of Haar basis onto the cone of nondecreasing functions and standardising them. In view of the above, we can say that the measure $q$ represents aggregated correlation coefficients pertinent to the copula $C$ and the selected set of functions. In econometrics, the indicators $\mathbb{1}(X\leq x_u)$ and $\mathbb{1}(Y\leq y_v)$ are labelled quantile hits. Evidently,
$$q(u,v)=\mathrm{Corr}[\mathbb{1}(X\leq x_u),\mathbb{1}(Y\leq y_v)]=\mathrm{Corr}[\mathbb{1}(X>x_u),\mathbb{1}(Y>y_v)].$$
Note also that
\begin{equation}\nonumber
q(u,v)=\int_0^1 \int_0^1 h_u(w)h_v(z) dC(w,z)
\end{equation}
and therefore,  from a mathematical point of view, the value of $q(u,v)$ is the Fourier coefficient of a density of $C$, if it exists, in the product system of  step functions $h_u(\cdot)$ and $h_v(\cdot)$. 

The properties of $q$ have been studied in \cite{Ledwina14arx,Ledwina15}  and recalled and discussed further in \cite{Cmiel}. In particular, $q(u,v) \in [-1,1]$ for all $(u,v) \in (0,1)^2$, $q$ is maximal (minimal) if and only if $Y=f(X)$ and $f$ is strictly increasing (decreasing) a.s. in the range of $X$, while $q \equiv 0$ if and only if $X$ and $Y$ are independent. For more details, refer to the papers mentioned just above.

In the following, we provide two new views on $q(u,v)$. The first relates $q$ to the pertinent least-squares regression line, while the second expresses $q$ via conditional probabilities. In order to describe new findings, for any fixed $(u,v)\in(0,1)^2$ we set $Z=(Z_1, Z_2)=(\mathbb{1}(X>x_u),\mathbb{1}(Y>y_v))$.
\begin{proposition}\label{i1}
For any $(u,v)\in(0,1)^2$, the best linear approximation of $Z_2$ by $Z_1$ has the form
$$\frac{z_2-(1-v)}{\sqrt{v(1-v)}}=q(u,v)\frac{z_1-(1-u)}{\sqrt{u(1-u)}}.$$
The slope of this line equals $P(Y>y_v|X>x_u)-P(Y>y_v|X\leq x_u)$.
Moreover, it holds that 
$$q(u,v)=\sqrt{\frac{u(1-u)}{v(1-v)}}\left[P(Y>y_v|X>x_u)-P(Y>y_v|X\leq x_u)\right]$$
$$\ \ \ \ \ \ \ \ =\sqrt{\frac{v(1-v)}{u(1-u)}}\left[P(X\leq x_u|Y\leq y_v)-P(X\leq x_u|Y> y_v)\right].$$
\end{proposition}
\begin{figure}[ht!]
\vspace{-1cm}
\centering{
\includegraphics[trim = 12mm 160mm 0mm 42mm, clip, scale=0.65]{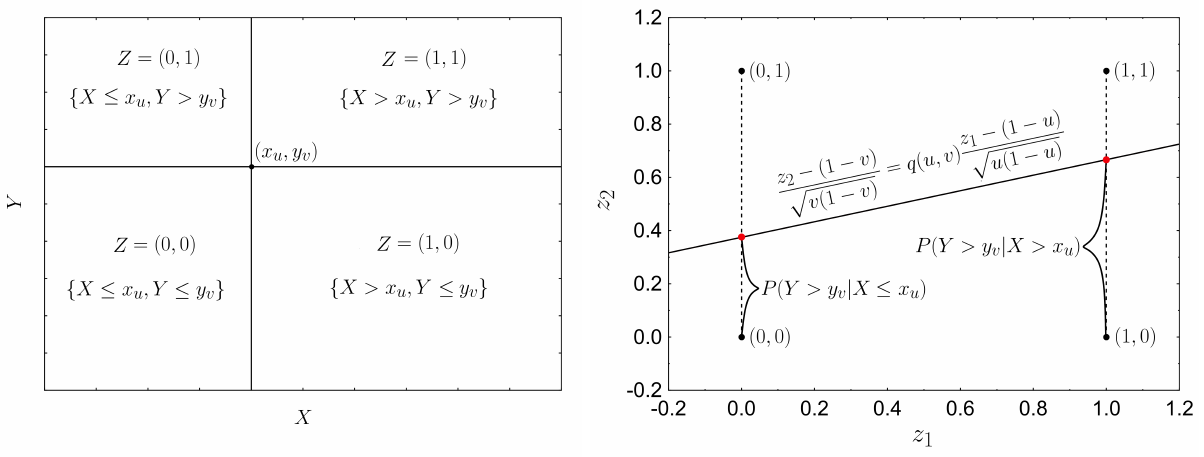}
}
\caption{Graphical presentation of Proposition \ref{i1} in an example where $q(u,v)>0$.}\label{I}
\end{figure}
Notice that $q(u,v)>0$ if and only if switching from the region $\{X\leq x_u\}$ to the region $\{X>x_u\}$ increases the probability of $\{Y>y_v\}$. The value of $q(u,v)$ reflects the intensity of this change. Figure \ref{I} summarises these findings. The standardisation of the slope, appearing in the above representation of $q(u,v)$, makes the empirical variant of $q(u,v)$ asymptotically $N(0,1)$ when $X$ and $Y$ are independent.     

In this paper, we would like to pay, among others, slightly more attention to one of the important properties of $q$ and its estimator which consists in preserving natural and popular order among copulas, namely {\it the concordance ordering}. The notion goes back to \cite{Yanagimoto}. For further extensions, throughout the discussion, and related references, see
\cite{Cambanis}. For two copulas $C_1$ and $C_2$, $C_2$ is more concordant than
$C_1$, in short $C_2 \succ C_1,$ means that large values of $X$ tend to be associated with larger values of $Y$ more frequently with $C_2$ than with $C_1$. Formally,  $C_2 \succ C_1$ means that $C_2(u,v) \geq C_1(u,v)$ for all $(u,v) \in [0,1]^2$. From (\ref{1}) it follows that stronger concordance implies larger $q$ and vice versa. Many known families of copulas, indexed by a single parameter, are ordered in such a sense. Bivariate normal and some elliptically contoured families, Gumbel-Hougaard, and Clayton families are prominent representatives. For more examples, see \cite{Nelsen}, pp. 136 - 137 and p. 188. A particular case where $C$ is more concordant than $\mathcal{I}$ is known under the label {\it positive quadrant dependence}, PQD in short. By the above, $q(u,v) \geq 0$ for all pairs $(u,v)$ if and only if $X$ and $Y$ are PQD. Similar conclusions in $q$ concern negatively ordered families of copulas. 

To conclude, note that in Section \ref{s6} we present a natural multivariate extension of $q$.

\section{Empirical checkerboard copula, new estimator of the measure $q$, and their properties}\label{s3}

Henceforth $(X_1,Y_1),...,(X_n,Y_n)$ are independent and identically distributed random vectors drawn from cdf $H, \;H \in {\cal H}^c$. By $F_n$, $G_n$ and $H_n$ we denote respective empirical cdf's in the $X$, $Y$ and $(X,Y)$ samples. Furthermore, $R_i$ is the rank of $X_i,\;i=1,...,n$, in sample $X_1,...,X_n$, while $S_i$ denotes the rank of $Y_i,\;i=1,...,n,$ in sample $Y_1,...,Y_n$.

We shall estimate $q$ using the plug-in method. Therefore, we start with a useful estimate of $C$. In this context note that in \cite{Ledwina14arx,Ledwina15} and \cite{Ledwina14} ordinary empirical copula $C_n$ was used to estimate $q$ and some limitations of this choice in related test construction were observed. In \cite{Cmiel} a new (symmetrised) estimator of $C$ was constructed and used for graphical analysis and to define stable test statistics. In the following, we propose applying the bilinear approximation of $C_n$, which yields nice properties of the resulting estimator $q$ and the related test statistic and is useful for graphical applications. 

\subsection{Bilinear estimator  $\bar C_n$ of $C$}\label{s3.1}

The standard estimate $C_n$ of the copula $C$ has the form
$$
C_n(u,v)=\frac{1}{n}\sum_{i=1}^n \mathbb{1}\left(\frac{R_i}{n} \leq u, \frac{S_i}{n} \leq v\right),\;\;\;(u,v) \in [0,1]^2.
$$

In this paper, we consider a smoothed variant of $C_n$, which is a bilinear interpolated version of $C_n$. To be specific, 
bilinear interpolation $\bar C_n$, say,  of $C_n$ at $(u,v) \in [0,1]^2$, expressed throughout the values of $C_n$ at the four points 
$\bigl(\frac{\lfloor nu \rfloor}{n}, \frac{\lfloor nv \rfloor}{n} \bigr)$, 
$\bigl(\frac{\lfloor nu \rfloor  +1}{n}, \frac{\lfloor nv \rfloor}{n} \bigr)$, 
$\bigl(\frac{\lfloor nu \rfloor}{n}, \frac{\lfloor nv \rfloor +1}{n} \bigr)$, 
and $\bigl(\frac{\lfloor nu \rfloor +1}{n}, \frac{\lfloor nv \rfloor +1}{n} \bigr)$,
where $\lfloor \bullet \rfloor$ stands for the integer part of $\bullet$, has the form
\begin{eqnarray}\label{3b} 
\bar{C}_n(u,v)&=&(1-nu + \lfloor nu \rfloor)(1-nv +\lfloor nv \rfloor)C_n\Big(\frac{\lfloor nu \rfloor}{n},\frac{\lfloor nv \rfloor}{n}\Big) + \nonumber\\ 
&&(nu-\lfloor nu \rfloor)(1-nv+\lfloor nv \rfloor)C_n\Big(\frac{\lfloor nu \rfloor+1}{n},\frac{\lfloor nv \rfloor}{n}\Big) +\nonumber \\
&&(1-nu +\lfloor nu \rfloor)(nv-\lfloor nv \rfloor)C_n\Big(\frac{\lfloor nu \rfloor}{n},\frac{\lfloor nv \rfloor+1}{n}\Big) +\nonumber \\
&&(nu-\lfloor nu \rfloor)(nv-\lfloor nv \rfloor)C_n\Big(\frac{\lfloor nu \rfloor+1}{n},\frac{\lfloor nv \rfloor+1}{n}\Big).
\end{eqnarray}
For every $u,v$ in [0,1], $\bar{C}_n(u,0)=\bar{C}_n(0,v)=0$ and $\bar{C}_n(u,1)=u,\;\bar{C}_n(1,v)=v$. Moreover, $\bar C_n(u,v)$ is a continuous function on $[0,1]^2$. 
The estimator $\bar{C}_n$ is the empirical checkerboard copula. For a comprehensive discussion of this class of estimators, see \cite{Genest17}. Hence, by their Remark 2, we infer that $\sup_{0 \leq u,v \leq 1} |C_n(u,v)-\bar C_n(u,v)| \leq 2/n$ almost surely.

Elementary calculations yield
$$
\bar{C}_n(u,v)=\frac{1}{n}\sum_{i=1}^n \Bigl\{\mathbb{1}(R_i \leq  \lfloor nu \rfloor) + (nu- \lfloor nu \rfloor)\mathbb{1}(R_i=\lfloor nu \rfloor +1)\Bigr\} \times
$$
\begin{equation}\label{6}
\Bigl\{\mathbb{1}(S_i \leq  \lfloor nv \rfloor) + (nv- \lfloor nv \rfloor)\mathbb{1}(S_i=\lfloor nv \rfloor +1)\Bigr\}.
\end{equation}
The formula (\ref{6}) shows that, given $(u,v) \in [0,1]^2$, $\bar{C}_n(u,v)$ is a linear rank statistic. Hence, under the independence of $X$ and $Y$, by \cite{Hajek}, pp. 61-62, we infer that
$$
{\text{E}}\bar{C}_n(u,v)=uv,\;\;\;{\text{Var}}\bar{C}_n(u,v)=(n-1)^{-1}{\{u(1-u)-\varepsilon_n(u)\}\{v(1-v)-\varepsilon_n(v)\}},
$$ 
where
$$
\varepsilon_n(t)=n^{-1}{\{nt-\lfloor nt\rfloor\}\{1-(nt-\lfloor nt\rfloor)\}}\in\bigl[0,{1}/{(4n)}\bigr],\ \ \ \ t\in[0,1].
$$

Below, we claim that  $\bar C_n$ preserves the concordance ordering. This result extends Proposition 1 in \cite{Ledwina14}, while its proof greatly simplifies an earlier argument patterned after \cite{Yanagimoto}.
\begin{proposition}\label{p1}
Suppose that the copula $C_{*}$ is more concordant than the copula $C_{**}$; $C_{*}\succ C_{**}$, in short. Then, under any fixed $(u,v) \in (0,1)^2$,  any $c \in \mathbb R$, and any $n$ it holds that
$$
P_{C_*}(\bar C_n(u,v) \geq c) \geq P_{C_{**}}(\bar C_n(u,v) \geq c),
$$
where $P_C$ denotes the probability law pertinent to the population copula $C$. Hence, if $C_{*}\succ C_{**}$ then, given $(u,v)$, $\bar C_n(u,v)$ is stochastically not smaller under $C_{*}$ than under $C_{**}$. 
\end{proposition}

\subsection{New estimator $\bar q_n$ of $q$}

 We shall consider throughout the estimator of $q$ given by
$$
\bar q_n(u,v)=\frac{\bar C_n(u,v) -uv}{\sqrt{uv(1-u)(1-v)}},\;\;\;(u,v) \in (0,1)^2.
$$
It has several useful properties, which we have summarised below.\\
\begin{proposition} For any $H \in {\cal H}^c$ we have
\begin{enumerate}
\item Since $\bar C_n$ is a {\it bona fide} copula, therefore $\bar q_n$ satisfies all properties of the quantile dependence function;
\item $\bar q_n$ is a continuous function on $(0,1)^2$ satisfying $-1\leq \bar q_n \leq 1$;
\item For each $v \in (0,1)\;\lim_{u \to 0} \bar q_n(u,v)=\lim_{u \to 1} \bar q_n(u,v)=0$;
\item For each $u \in (0,1)\;\lim_{v \to 0} \bar q_n(u,v)=\lim_{v \to 1} \bar q_n(u,v)=0$;
\item In view of Proposition \ref{p1}, $\bar q_n$ preserves concordance ordering. This means that for any two copulas $C_{*}$ and $C_{**}$, such that $C_{*}\succ C_{**}$, and any fixed $(u,v)\in (0,1)^2$, $\bar q_n(u,v)$ is stochastically not smaller under $C_{*}$ than under $C_{**}$. In such a sense, $\bar q_n$ quantifies this notion;
\item By Section \ref{s3.1}, under independence of $X$ and $Y$, the estimator $ \bar q_n$ has mean 0 and variance close to $1/n$. In addition, we infer that for each $(u,v) \in (0,1)^2$, under the independence of $X$ and $Y$, $\sqrt n \bar q_n(u,v)$ is asymptotically N(0,1). 
\end{enumerate}
\end{proposition}
\begin{remark}
The above definition of $\bar q_n$ can be naturally extended to multivariate distributions with continuous margins.  For details, see Section \ref{s6}. The generalisation is related to some results by \cite{Deheuvels_80} on piecewise linear multivariate empirical copulas. Moreover, in view of the developments by \cite{Genest17}, a similar extension is possible for the multivariate case with arbitrary univariate margins.   
\end{remark}
Throughout $\bar m$ and $\bar M$ denote the minimal and maximal value of $\bar q_n$ over $(0,1)^2$, respectively. By the properties of $\bar q_n$, we have $-1 \leq \bar m \leq \bar M \leq 1.$

\subsection{Illustration}

\begin{figure}[ht!]
\centering{
\includegraphics[trim = 10mm 50mm 10mm 30mm, clip, scale=0.75]{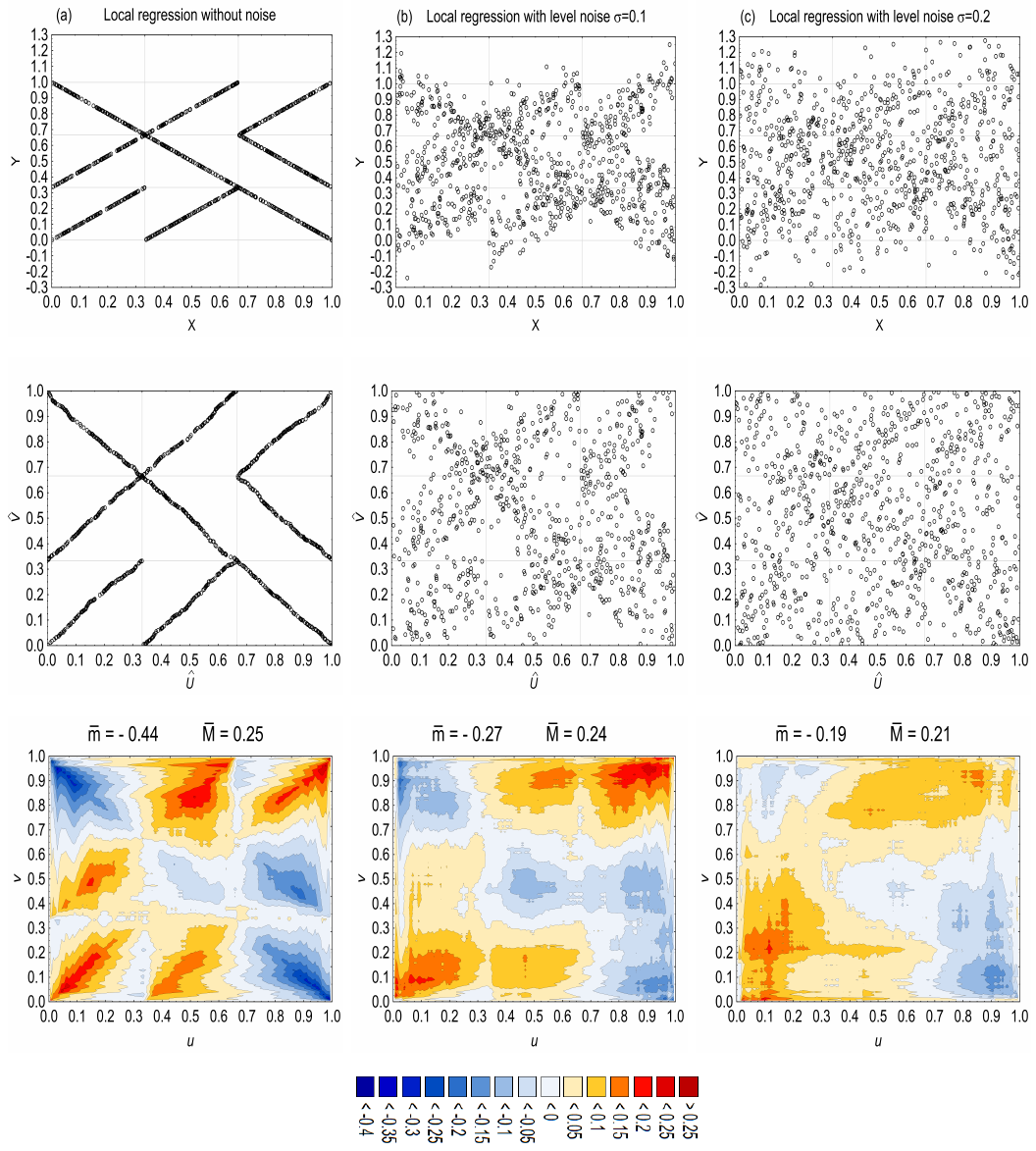}
}
\caption{Empirical quantile dependence function $\bar q_n$ for locally linear regression against level of noise: (a) Setting with  no noise; (b) Setting with $\sigma =0.1$; (c) Setting with $\sigma=0.2.$}\label{F2}
\end{figure}

Below, we illustrate how $\bar q_n$ reacts to the growing level of noise in an artificial locally linear regression model.  Pertinent realisations are shown in the first row in Figure \ref{F2}. Let us start from the left top panel. The square $[0,1]^2$ is divided into 9 squares of equal size. For each of the 9 selected squares, we generate 100 observations $(X,Y)$ defined as follows: $X$ is uniformly distributed over the horizontal side of the selected rectangle, while $Y$ is a linear function of $X$ - one of the diagonals of the rectangle. In the upper left panel, no noise is added to such data, while in the middle and upper right panels to each observation independent $N(0,\sigma^2)$ distributed noise is added, with $\sigma=0.1$ and $\sigma=0.2$, respectively. Since our estimate is based on ranks $(R_i,S_i),\; i=1,...,n,$ therefore, in the next row we also present scatter plots pertinent to pseudo-observations $(\hat U_i, \hat V_i)$, where $\hat U_i=R_i/n$ and $\hat V_i =S_i/n$. The bottom row in Figure \ref{F2} presents the related shapes of $\bar q_n(u; v)$. Recall also that, by Proposition 3, $\bar q_n$ inherits all properties after $q$.

\newpage

Figure \ref{F2} shows that the proposed estimator
effectively visualises the complex dependence structure, highlighting areas of positive
and negative dependence. For example, consider the square $(0, 0.3]^2$ in column (a). The points marked in red represent strongly positive values of the estimator. This implies that locally, for related arguments $(u,v)$, the empirical copula $\bar C_n(u,v)$ is much larger than could be expected under the independence hypothesis. In other words, in this region, relatively small pseudo-observations $\hat U_i$'s tend to associate relatively small $\hat V_i$'s much more often than expected. For other points in $(0,0.3]^2$ the estimator $\bar q_n$ is close to 0. The above clearly suggests a strong positive association of the  pseudo-observations  in this region. Simultaneously, due to Proposition 1, a similar conclusion is valid for the original data. Namely, for values of $X$ and $Y$ smaller than their $x_u$ and $y_v$ quantiles of order at most 0.3, with $(u,v)$ located along and near the diagonal of $(0,0.3]^2$, the chance of small values of $X$ to be associated with small values $Y$ is noticeably higher than the alternative tendency. A similar interpretation applies to the two remaining cases.

The discussed exemplary relationship, exhibited by $\bar q_n$,  is  best visible in both  plots of observations in the local regression model without noise. For noisy data, as the level of noise increases, this tendency, as well as other relations displayed in the two successive panels in (a), become  less transparent. In particular, the middle panel in (c) resembles to a large extent realizations of uniform distribution on $(0,1)^2$, while the estimator $\bar q_n$ still reports some areas of relatively strong dependence under the pertinent level of noise. Thus, in practice, the estimator $\bar q_n$ can be used to detect such local relationships even when they are not visible in the scatter plot, for example. Of course, if the local dependence is weak, the  estimator $\bar q_n$ may not capture it  for a small sample size. To determine whether the values of the function $\bar q_n$  are locally considerably  far from zero, i.e., to conclude that the local dependence is significant, a local statistical test is needed. This is exactly what the acceptance regions and the dependence diagram, described in the next section, do.

\section{Inspection of dependence structure and first illustrations}\label{s4}

In recent years, we observed growing interest in a deeper understanding of some questions related to testing statistical hypotheses. Examples of challenging goals posed in this context include the interpretation of global relations between the postulated model and the empirical one upon rejection of the null hypothesis and identification of local discrepancies. For a discussion of these and other similar-nature inferential targets, see \cite{Cazals}, \cite{Ma}, \cite{Zhang}, \cite{Algeri}, \cite{Dvorak},  \cite{Ducharme}, \cite{Ledwina24}, \cite{Jimenez}, and references therein. In addition, note that understanding the character of the leading discrepancies between the postulated model and the data is also important to decide on an adequate test statistic to assert the validity of the null hypothesis. For a related discussion, see \cite{Ducharme}.

We shall discuss below the two of the above mentioned questions, when posed in the case of testing and identifying dependency nature. 

The first problem, which we shall consider,  is inspection of the local dependence structure via acceptance regions, which is our new impact on the area. 
Our approach is presented in Section \ref{s4.1}. The solution relies on local quantification of the magnitude of the estimator $\bar q_n$ of the quantile dependence function. As we shall see, the approach provides clear hints on subsets of pairs of marginal quantiles that exhibit regions of highly unlikely behaviour of the estimate $\bar q_n$ in the case where the independence hypothesis would be true. Hence, this information also has a high potential impact to account for the rejection of the global independence relation. The last mentioned problem shall be treated in Section \ref{s5}. There we propose a new powerful test, pertinent to the new estimator $\bar q_n$, as well.  

The second question, which we shall briefly present in Section \ref{s4.2z}, is the \cite{Zhang} approach to evaluating the global relationship between two continuous random variables. We provide more comments on this approach in Appendix A.5. In Sections \ref{s4.2} and \ref{s5.3} we apply both approaches to some sets of observations and discuss the result. It is worth noting that Zhang's approach has been recently applied by \cite{Xiang} to an impressive study of dependencies in genomic data.

\subsection{Discretization, simultaneous acceptance regions, and the dependence diagram}\label{s4.1}

Having defined the smoothed variant $\bar C_n$ of the ordinary estimator $C_n$ of a copula $C$ let us introduce the related weighted rank process
\begin{equation}\label{7}
\bar Q_n(u,v)=\sqrt n \bar q_n(u,v)=\sqrt n \frac{\bar C_n(u,v) -uv}{\sqrt{uv(1-u)(1-v)}},\;\;\;(u,v) \in (0,1)^2.
\end{equation}
The standard next step would be to consider some statistics being functionals of this continuous time process. However, we shall proceed in a less standard way and restrict our attention to some selected finite-dimensional realisations of this process, which are much easier to handle. The grid-based approach is not uncommon in practice, especially in econometrics, biometrics, and signal processing. The contribution of \cite{Reshef11} has raised great interest in grid-based approaches to independence testing and stimulated a lot of developments; see \cite{Reshef18} for related discussion. A recent paper by \cite{Gonzalez} provides updated comments and an implementation of such an approach when detecting independence from the perspective of signal processing. The independence testing approaches elaborated by \cite{Ma} as well as by \cite{Zhang} are also based on adequately defined discretizations. As nicely phrased by \cite{Ma}, p. 247: ``{ In practice, one is typically only interested in the dependency up to a practical level of details}''. Both the above-mentioned papers also provide and discuss some formalizations of such approximate independence. The fundamental difference between their and our approaches to discretization is that they refer to standard cells, i.e., a comparison of the joint density in $[0,1]^2$ and the product of marginal ones, while we follow the contribution by \cite{Ledwina14}, and we rely on the comparison of the joint cdf in $[0,1]^2$, i.e., the copula in our setting and the product of the respective marginals. Compared with the above-mentioned, this solution allows for finer discretization, given the sample size $n$, as it is based on cumulative sums of probability mass, not cell frequencies, and removes, to a large extent, problems with too few observations in standard cells. Note also that in a recent paper by \cite{Dvorak} one of their tests construction starts from the empirical bivariate cdf on ${\mathbb R}^2$ evaluated on a finite grid of points on ${\mathbb R}^2$.

The grid, which we shall use, follows from the diadic partition of [0,1]. More precisely, given $s,\;s=0,1,...$ , set $d(s)=2^{s+1}-1$ and consider points of the form
$$
p_{s,j}=\frac{j}{2^{s+1}}=\frac{j}{1+d(s)},\;\;\;j=1,...,d(s).
$$
For given sample size $n$, we shall restrict our attention to $s=s(n)$, where $s(n)$ is a user-defined increasing sequence, while $j=1,...,d(n)$, where $d(n)=d(s(n))$. The related inspection points have the form $(p_{s(n),j},p_{s(n),j'}),\;j,j'=1,...,d(n)$. Consequently, we shall consider the values $\bar{Q}_n(p_{s(n),j},p_{s(n),j'}),\;j,j'=1,...,d(n)$, of the rank process (\ref{7}). 

Now, let us turn back to the formula (\ref{2}), the form of functions $h_u(\cdot)$ and $h_v(\cdot)$ defined there, and set $u=p_{s(n),j}$ and $v=p_{s(n),j'}$. With this notation, we get the following.
\begin{equation}\label{8}
\bar Q_n(p_{s(n),j},p_{s(n),j'})=\sqrt n {\;{W}}_{j,j'}^o,\ \mbox{where}\;{{W}}_{j,j'}^o=\hspace{-0.1cm} \int_0^1\hspace{-0.1cm} \int_0^1 \hspace{-0.1cm}h_{p_{s(n),j}}(w) \times  h_{p_{s(n),j'}}(z) d\bar C_n(w,z).
\end{equation}
In other words, $\bar Q_n(p_{s(n),j},p_{s(n),j'})$ is standardised by $\sqrt{n}$, empirical Fourier coefficient relevant to $\bar C_n$ and the product function $h_{p_{s(n),j}}(\cdot) \times h_{p_{s(n),j'}}(\cdot)\;$. This product system was already used by \cite{Ledwina14} in the context of testing for positive quadrant dependence. 

To identify the actual nature of the dependence, we study the magnitude of these selected empirical Fourier coefficients. We propose to restrict attention to successive deciles in marginal cdf's. Such `accuracy' is standard choice in the econometric literature, for example. Obviously, when some very sophisticated dependence structures are under consideration, then adequately fine accuracy would be appropriate. The next steps are as follows.
\begin{itemize}
\item Define ten intervals $I_1=(0,0.1], I_2=(0.1,0.2],...,I_{10}=(0.9,1.0)$ and their products $\Pi_{k,l}=I_k \times I_l,\;k,l=1,...,10.$
\item Then, given $n$, $s(n)$ and $d(n)=d(s(n))$, for each fixed $k$ and $l$, find subsets, $P_{k,l}$, say,  of points $(p_{s(n),j},p_{s(n),j'}),\;j,j'=1,...,d(n),$ that fall into the cell $\Pi_{k,l}.$ We skip in the notation the dependence of $P_{k,l}$ on $n$.
\item Now, set
$$
\hspace{-0.5cm} L^{-}(\bar Q_n, \Pi_{k,l})=\hspace{-0.2cm}\min_{j,j' \in P_{k,l}} \hspace{-0.1cm} \bar Q_n(p_{s(n),j},p_{s(n),j'})\;\mbox{and}\;
L^{+}(\bar Q_n, \Pi_{k,l})=\hspace{-0.2cm}\max_{j,j' \in P_{k,l}} \hspace{-0.1cm} \bar Q_n(p_{s(n),j},p_{s(n),j'}). 
$$
for local minima and maxima of the discretised variant of $\bar Q_n(u,v)$ on the selected set of indices.
\item  Under the independence hypothesis, given $\alpha$, the cell $\Pi_{k,l}$, and the process $\bar Q_n$, find two barriers $\ell^{-}(n,\alpha,\Pi_{k,l})$ and 
$\ell^{+}(n,\alpha,\Pi_{k,l})$ defined as follows
\begin{equation}\label{9}
P(L^{-}(\bar Q_n, \Pi_{k,l}) \geq \sqrt n \ell^{-}(n,\alpha,\Pi_{k,l}) ) \geq 1-\alpha,
\end{equation}

\begin{equation}\label{10}
P(L^{+}(\bar Q_n, \Pi_{k,l}) \leq \sqrt n \ell^{+}(n,\alpha,\Pi_{k,l}) ) \geq 1-\alpha.
\end{equation}
\end{itemize}

Following the terminology introduced in \cite{Ducharme}, in the case of goodness-of-fit testing, we call (\ref{9}) and (\ref{10}) simultaneous one-sided acceptance regions at the level $1-\alpha$. Since the rank process $\bar Q_n$ is distribution-free under the independence hypothesis, therefore ordinary Monte Carlo is sufficient to calculate the barriers $\ell^{-}(n,\alpha,\Pi_{k,l})$ and $\ell^{+}(n,\alpha,\Pi_{k,l})$. 

From (\ref{9}) i (\ref{10}) it follows that
$$
P\bigl(\sqrt n \ell^{-}(n,\alpha, \Pi_{k,l}) \leq L^{-}(\bar Q_n,\Pi_{k,l}), \;L^{+}(\bar Q_n,\Pi_{k,l}) \leq \sqrt n \ell^{+}(n,\alpha,\Pi_{k,l})\bigr) \geq 1-2\alpha.
$$
Hence, $\bigl[\ell^{-}(n,\alpha/2, \Pi_{k,l}),\;\ell^{+}(n,\alpha/2,\Pi_{k,l})\bigr]$ defines the local acceptance region at the level $1-\alpha$ for the arguments of the function $q$ belonging to $\Pi_{k,l}$. 

The interpretation of (\ref{9}) and (\ref{10}) is clear. For instance, if the selected local minimum is below the pertinent barrier, this indicates that the data seem incompatible with independence in the considered region, as at least one related empirical Fourier coefficient is too small. A similar conclusion applies to local maxima. Given the data and the acceptance level $\alpha$, we can check for each of $\Pi_{k,l},\;k,l=1,...10,$ whether $\bar q_n(p_{s(n),j},p_{s(n),j'}),\;j,j' \in P_{k,l}$, belongs to $[\ell^{-}(n,\alpha/2,\Pi_{k,l}),\ell^{+}(n,\alpha/2,\Pi_{k,l})]$, lies below $\ell^{-}(n,\alpha/2,\Pi_{k,l})$, or lies above $\ell^{+}(n,\alpha/2,\Pi_{k,l})$. Throughout the paper, we have used white, blue, and pink colours, respectively, to mark in each $\Pi_{k,l}$ the actual situation. Such a display, 
we shall call \textit{the dependence diagram}.  

It is evident that any other discretization can be applied in the above construction whenever convenient. 

In Sections \ref{s4.2}, \ref{s5.3}  we present some real-life applications of the dependence diagram. Appendix A.3 briefly discusses an influence of the value of $d(n)$ on the magnitude of barriers. In our paper, we have considered moderately large sample sizes and related moderately large grid dimensions. In the case where one has a very large sample and likes to consider a very fine grid, it could be useful to modify the above procedure replacing $L^{-}(Q_n,\Pi_{k,l})$
and $L^{+}(Q_n,\Pi_{k,l})$ by their counterparts based on statistics like $T_n$, introduced in Section \ref{s5}, below.

We conclude here with  preliminary remarks on an alternative way to detect a dependence structure.

\subsection{Preliminary comparison of the approach of Zhang's (2019) to ours}\label{s4.2z}

Our approach to identify significant local deviations from independence is by inspecting the local minima and maxima of the discretised variant of $\bar q_n$. In other words, due to one of the interpretations of $\bar q_n$, we investigate the minima and maxima of selected empirical Fourier coefficients (\ref{8}). The different formulated goals and the other scope were defined in \cite{Zhang}. He aimed to find some general pattern of mass allocation after the rejection of independence. In consequence, his approach relies on proposing a kind of pattern book and verifying which template fits the data in the best way. In the original Zhang contribution, in its extension \cite{Lee}, and in the applications in \cite{Xiang}, the attention is restricted to nine patterns depicted by blue and white rectangles in Figure 2 of \cite{Xiang}. The procedure designed for the selection of a pattern is denoted by Max BET. For comparison with our approach, in Sections \ref{s4.2}, \ref{s5.3}, and A.5 we shall also use these nine patterns. 
Note that BET in Zhang's theory is an abbreviation of Binary Expansion Testing. In Appendix A.5 we present new point of view on description and selection of Zhang's patterns and present Max BET in detail. In particular, we notice there that Zhang's approach simply exploits empirical Fourier coefficients $W_{i,j}$, say, pertinent to $C_n$ and the product system of Walsh functions.

\subsection{Graphical analysis of two popular data sets}\label{s4.2}

In this section, we shall illustrate how our approach to visualisation and evaluation of dependency structure works in the context of two known sets of data. We also compare our findings with the results of Zhang's pattern fitting. In Section \ref{s5.3} additional sets of data are analysed, which refer to more complex dependence structures than those in this section.

Recall that we assumed throughout that the joint cdf $H$ belongs to ${\cal H}^c$. Therefore, no ties are expected. When in our examples a number of ties appears due to some rounding of the measurements, then we apply randomised ranks $R_1^*,...,R_n^*$ and $S_1^*,...,S_n^*$ and keep the name pseudo-observations for the pairs $(R_i^*/n,S_i^*/n),\;i=1,...,n.$ 

Here we consider {\it ethanol measurements}, $n=88$, described in \cite{Simonoff},  and a subset of {\it Danish fire insurance data}, $n=517$ pairs of contents and profits, as already analysed by \cite{Gijbels}. There are 5 pairs of tied observations among 88 considered in our first example, and there are no ties among 517 observations in the second case. 

\begin{figure}[ht!]
\hspace{1cm}
\centering{
\includegraphics[trim = 20mm 115mm 10mm 20mm, clip, scale=0.7]{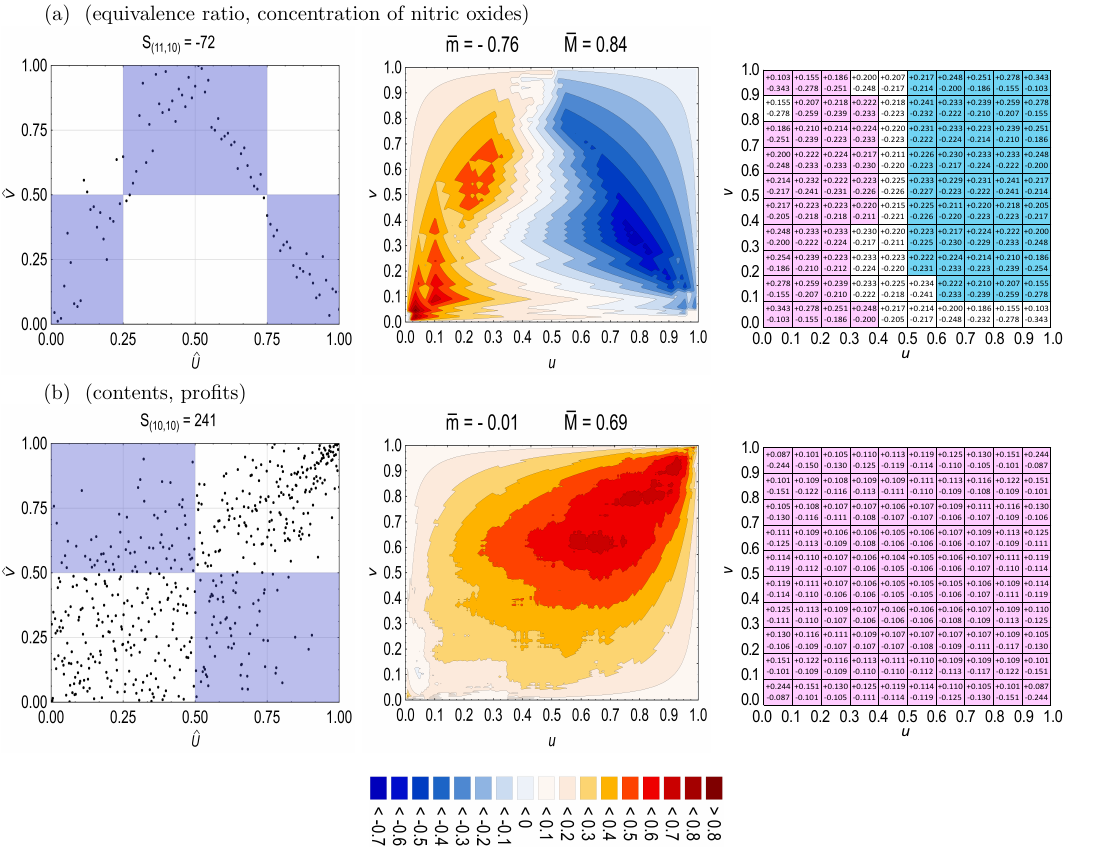}
}
\caption{Scatter plot and Zhang's pattern, estimated $q$, and the dependence diagram: (a) ethanol data; (b) Danish fire insurance data (contents and profits). Barriers in dependence diagrams are based on 100 000 MC runs, $d(88)=31$ for ethanol data and $d(517)=255$ for insurance data.} \label{F3}
\end{figure}

We start with some graphical output for the two sets. In the first column in Figure \ref{F3} we present scatter plots of related pseudo-observations along with patterns fitted by Zhang's procedure and the values of pertinent Zhang's symmetry statistics. In the next two columns, we show estimated quantile dependence functions $\bar q_n$ and pertinent acceptance regions for related subsets of bars. When drawing dependence diagrams, we consider the $31 \times 31$ grid in the first example and the $255 \times 255$ grid in the second. The resulting barriers are shown on a larger scale in Appendix A.3. Their stability against $d(n)$ is studied there as well. More comments on the data are given below.

The ethanol data relate the concentration of nitric oxides in engine exhaust to the equivalence ratio, which is a measure of the richness of the air-ethanol mix from burning ethanol in the engine. This is a widely studied set of observations. These data were provided as support material for the book by \cite{Simonoff}. They are
available at {\tt https://pages.stern.nyu.edu/$\sim$\\jsimonof/SmoothMeth/Data/ASCII/allfiles}. On p. 137 of the book, in Fig. 5.3, we see a scatter plot of the original 88 data points and can read following their description: ``{ The curve shows an increase in nitric oxides up to an equivalence ratio of about 1, followed by a steady decrease. This pattern might suggest a parabolic (quadratic) relationship, but the superimposed quadratic least-squares fit shows that this is not the case.}'' For more complex regression models for these data, see \cite{Hurn} and \cite{Young}, for example. 

The data were analysed in \cite{Kallenberg} and their test definitively rejected independence. They have been reconsidered by \cite{Fisher01}, who changed the axes, applied to such rotated pairs of related pairs of pseudo-observations  their $\chi$-plot (\cite{Fisher85}) and concluded that:  ``{This example hardly calls for use of the $\chi$-plot... the values in the plot are more dispersed than one would expect if the variates were independent''}. \cite{Genest03} have supplemented these findings, using their K-plot, as follows:  ``{The K-plot, provided as a complement, suggests the presence of mild negative association in the data''}. 

Our application of empirical $q$ provides much clearer and quantified information on the association of the two variables, which clearly supports the two-phase dependence structure of the data. Left to the $0.4$ quantile of the equivalence ratio, we notice a strong positive association, while right to the median of the equivalence ratio, the plot exhibits a strong negative dependence; see the middle panel in (a). Simultaneous $90\%$ local acceptance regions confirm these findings and indicate quantile regions in which deviations from independence (measured at the grid points) are locally significant positive/negative at the nominal level 0.05. 

Let us give some less formal comments on our analysis of these data. The display of $\bar q_n$ and the related values of  $\bar m$ and $\bar M$ clearly support the conclusion of \cite{Fisher01} saying that it is simply not credible for the observations to be independent in this experiment. In spite of this general conclusion, one can be interested in a more specific question: which levels of the equivalence ratio are really responsible for an increase/decrease of the concentration of nitric oxides. To get some evidence of this kind, our one-sided acceptance regions provide a support. Namely, they define local one-sided tests (pertinent to the values of $\bar q_n$ in related grid points) at a prescribed significance level. In our application, we used the 0.05 level for each of the one-sided tests. These tests provide information, for each decile$\times$decile region $\Pi_{k,l},\; k,l=1,...,10$, separately, if the observed large positive/small negative deviations from independence are statistically significant. Next, this information is summarised graphically in the dependence diagram, in which the regions $\Pi_{k,l}$ with significant positive/negative deviations are marked in red/blue, respectively, while the remaining cases are displayed in white colour. To be specific, consider, for example, three regions $\Pi_{1,5}$, $\Pi_{5,5}$, and $\Pi_{10,5}$. In $\Pi_{1,5}$ the dependence diagram is read. This means that in this region the statistic 
$L^{+}(\bar Q_n, \Pi_{1,5})$ is significantly large and indicates the existence of significantly large positive local deviations from the independence (at the sampling points from the grid in this region). Similarly, in $\Pi_{10,5}$ the dependence diagram is blue, as in this region $L^{-}(\bar Q_n, \Pi_{10,5})$ is significantly small, negative. Finally, no significant deviations from independence were observed in the grid points of $\Pi_{5,5}$. Hence, this square is white. The dependence diagram shows that the concentration of nitric oxides, in each of the ten quantile ranges, is not significantly related to the equivalence ratio in the central range of quantiles (0.4,0.5] of this factor. In contrast, right to the median of the explanatory variable, statistically significant negative influence of this factor to most of levels of nitric oxides, observed in the plot of $\bar q_n$, is confirmed.

Zhang's approach yields the largest asymmetry represented by $S_{(11,10)}=-72$. The negative value of the selected symmetry statistic implies that the sum of blue rectangles is the identified dependence structure. The related $p$-value is equal to 0. Evidently, approximately parabolic curve indeed fits to this template.  However, other patterns could also be fitted. We briefly discuss this point in Appendix A.5. Therefore, we claim that our approach is more flexible and accurate.

Danish fire insurance data are well-known datasets widely used in extreme value analysis. It was first made publicly available by \cite{McNeil}. For further analysis and related literature, see \cite{Embrechts} and \cite{Haug}. In fact, these data are available from R package {\it fitdistrplus}. The set collects three types of claim referring to building loss (B), content loss (C) and profit loss (P) and consists of 2 167 claims.   We follow the exposition in \cite{Gijbels} and consider only positive claims in all three losses. This reduces the sample size to $n=517$. \cite{Gijbels} investigated whether the three pairs (B,C), (B,P), and (C,P) follow a positive quadrant dependence (PQD) structure. For an illustration, we present here our findings for the (C,P) pair. 

The first impression from the scatter plot of (C,P), presented in the first panel of (b) in Fig. \ref{F3}, is the visible concentration of the pseudo-observations towards the upper-right corner. Some linear trend (along the main diagonal) can also be noticed. Zhang's approach selects the sum of presented white regions, reports a very large asymmetry for this template, and provides a related $p$-value equal to 0. In turn, the estimated $q$ is positive on $(0,1)^2$, clearly exhibits the underlying specific dependence structure, and thus supports our observations on the scatter plot. It is also worth noting that the display of $\bar q_n$ is much better readable than the level plot (i) in Fig. 5 of \cite{Gijbels}. The decisive role of the scaling factor $\sqrt{uv(1-u)(1-v)}$ is clearly manifested in this case. Our dependence diagram returns the local positive dependence, at the acceptance level of 0.95, in all regions defined by marginal deciles and the grid of $255 \times 255$ points. This strongly supports the PQD structure for these data, and, at this point, the conclusion is in agreement with the findings of \cite{Gijbels}. Again, we see that our approach has provided more careful insight than the Zhang template.

\section{New test of global independence}\label{s5}

\subsection{Test statistic and its properties}

\begin{figure}[ht!]
\centering{
\includegraphics[trim = 28mm 195mm 40mm 30mm, clip, scale=0.85]{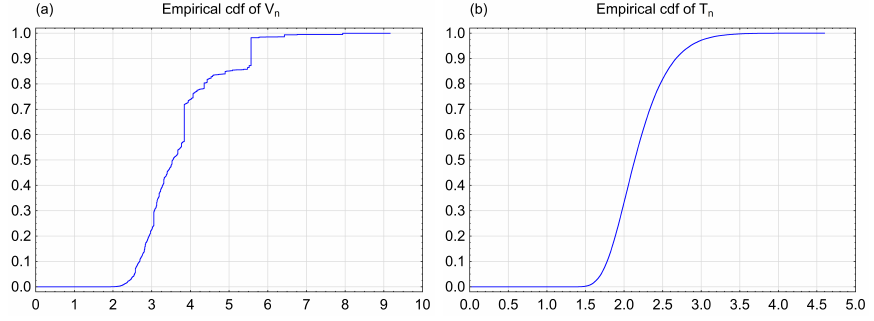}
}

\caption{Plots of the empirical distribution functions of $V_n$ and $T_n$ under ${\mathbb H}_0$, $n=128$, and 100 000  MC runs; $d(n)=63$, $t(n)=0.95$.}\label{F4}
\end{figure}

Recall that we consider $(X,Y)$ with continuous marginals $F$ and $G$ and a joint cdf $H(x,y)$. We would like to test
$$
{\mathbb H}_0 : H(x,y)=F(x)G(y)\;\;\;\mbox{for all}\;\;\;(x,y) \in {\mathbb R}^2.
$$
We shall focus our attention on the test statistic for ${\mathbb H}_0$ that reflects the global maximum of $\bar Q_n = \sqrt n \bar q_n$ over the grid pertinent to the $[d(n)]^2$ partition points. The simplest solution would be to consider
\begin{equation}\label{12}
V_n=V_n(d(n)) = \max_{1 \leq j,j' \leq d(n)} \Bigl |\bar Q_n\bigl(p_{s(n),j},p_{s(n),j'}\bigr)\Bigr |.
\end{equation}
However, it can be seen that for moderately large sample sizes and moderately large $d(n)$'s, under ${\mathbb H}_0$, the maximum in (\ref{12}), in a considerably large part of cases, is reached at the vertices of the bivariate grid. In addition, a relatively high concentration of arguments of (\ref{12}) is observed near the edges of the grid. This results in relatively large critical values. Moreover, the empirical cdf of $V_n$ is, in principle, discrete, which strongly affects the resulting empirical quantiles. Therefore, to eliminate these drawbacks, we sort the $[d(n)]^2$ values of $|\bar Q_n\bigl(p_{s(n),j},p_{s(n),j'}\bigr)|$ according to the increasing magnitude. We denote the resulting order statistics in short  by $\bar{ |Q|}_{(1)} \leq ... \leq \bar {|Q|}_{(K(n))} $, where $K(n)=[d(n)]^2$.
Next, we take $t(n) \in (0,1)$, set $\kappa (n)=\lceil  t(n) K(n) \rceil$, where $\lceil \bullet \rceil$ stands for the smallest integer not smaller than $\bullet$, and replace the global maximum (\ref{12}) by the following statistic
\begin{equation}\nonumber
T_n = T_n (d(n),t(n))=\frac{1}{K(n) - \kappa (n) +1} \sum_{k= \kappa(n)}^ {K(n)} \bar {|Q|}_{(k)}.
\end{equation}
This estimate of the maximum of $q$ is much more stable than $V_n$;  under ${\mathbb H}_0$, in particular. In our simulations of powers, presented in Sections \ref{s5.2} and A.1, we took $t(n)=0.95$. For an illustration, with this choice of $t(n)$, for $n=128$, which we use in simulations of powers, and $\alpha=0.10, 0.05, 0.01$, 100 000 MC simulation runs yield the following critical values of $V_n$ : 5.57, 5.57, 6.43, respectively, while for $T_n$ we have got 2.68, 2.86 and 3.24. Figure \ref{F4} presents empirical cdf's of both statistics under ${\mathbb H}_0$. For more detailed information about the above mentioned issues see Appendix A.2.
\newpage

\begin{theorem}\label{t1} Suppose that
\begin{enumerate}
\item $\frac{\partial}{\partial u}C(u,v)$ and $\frac{\partial}{\partial v}C(u,v)$ exist and are continuous on  $(0,1)\times[0,1]$ and $[0,1]\times(0,1)$, respectively;
\item $t(n) \in (0,1),\ d(n) \rightarrow \infty$ and $d(n)/\sqrt{n}\rightarrow 0$ when $n \rightarrow \infty$. 
\end{enumerate}
Then the test that rejects ${\mathbb H}_0$ for large values of $T_n$ is consistent.
\end{theorem}
\noindent
\begin{remark}
Note that in \cite{Cmiel} another consistent sup-type test of independence, pertinent to $q$, has been constructed; cf. $\mathcal{D}^*_{\kappa,s,n}$ in (13) therein. The construction exploited another estimate of $q$ and another smoothing of the sup-type statistic, which was based on the pertinent weighted empirical rank process. 

The new solution $T_n$ is simpler, more natural and more powerful under finite sample sizes than $\mathcal{D}^*_{\kappa,s,n}$ studied in \cite{Cmiel}. The new test is directly related to the discretization introduced in the present paper and allows for immediate extension to higher dimensions. Note also that $t(n)$ can be any sequence with values in the interval (0,1). It defines, in a certain way, a norm of discretized $\bar Q_n$ which for $t(n)<1/K(n)$ is equivalent to the $l_1$ norm and for $t(n)>(K(n)-1)/K(n)$ coincides with $l_{\infty}$ the norm. For further comments, see the final paragraphs of the next section. Observe also that Assumption 2 of Theorem \ref{t1} calls for some caution when deciding on a range of $d(n)$. However, our simulations show that $d(n)$ close to $n/2$, roughly speaking, yields satisfactory finite sample results. Appendix A.2 provides some evidence and discussion on the role of $d(n)$ and $t(n)$ as well as some recommendations on their choice in practice.
\end{remark}

\subsection{Power study - illustrative cases}\label{s5.2}

We carried out a simulation study to evaluate the performance of $T_n$ and to compare it with several state-of-the-art tests. Other methods we tested include rank version of distance covariance dCov (\cite{Szekely}), copula-based kernel test $\hat{D}_C(X,Y)$, DC in short, (\cite{Reddi13}), the HHG test (\cite{Heller13}), sign covariance $\tau_n^*$ (\cite{Bergsma}), Max BET with $d_{max}=4$ (\cite{Zhang}), and its improvement - the Ensemble BET test with $d_{max}=4$ (\cite{Lee}). We have also considered the minimum $p$-value of HHG and $T_n$, denoted by HHG$/T_n$. Here we only discuss and summarise this experiment, while more details are postponed to Appendix A.1. 

We treat the test statistic HHG as a benchmark as it is one of the best solutions when not very sophisticated deviations from independence are expected. For detailed evidence, see \cite{Heller13,Heller16}, \cite{Reshef18}, and \cite{Cmiel}. 

To avoid misunderstanding, note that in \cite{Zhang} two different statistics are introduced under the same label Max BET. First serves to select a template, cf. our Section \ref{s4.2z}, while the second one is better adjusted to verify global independence hypothesis. The last is described in Section 4.5 in \cite{Zhang}. In our power study here and in Appendix A.1 we use the second mentioned Zhang statistic. 

We start with some general comments on the simulation scenario. Many models used in recent simulations have the structure $(X,f(X)+\epsilon),$ where $X$ has a uniform distribution on [0,1], $\epsilon \sim {\text N}(0,\sigma^2)$, while $f$ is known. This way of modelling alternatives looks attractive, at least at first glance. However, without doubt many important models considered, e.g. in medical research, economics, and quantitative finance, are not included. \cite{Reshef18}, p. 152, recommend paying attention to local dependence, \cite{Thas} advocate reflecting realistic situations, \cite{Vexler17} emphasise the importance of random effects, among others. In turn, in insurance and mathematical finance, distributions with heavy tails play an important role. For a general discussion on the last mentioned point, see \cite{Cont}. For an illustration, in case of analysis of some risk measures, see \cite{Mainik}. We also refer to \cite{Genest19} for their simulation scenario.

In any case, to form a differentiated list of examples, some way of comparing successive models is welcome. The first warning is that analytical forms of the relations, such as joint densities or functional relationships, can sometimes be misleading, as they can provide in fact very similar dependence structures. The point is that many testing procedures are based on ranks and hence forms of pertinent copulas play an essential role, not the original, nicely looking, analytical formulas. \cite{Kallenberg} have provided presumably the first attempt to understand the empirical powers of some tests of independence in relation to the structure of copula models for alternatives. The goal was achieved  by calculating and displaying, for some standard bivariate models,  a set of  first Fourier coefficients, in the product Legendre's system, of pertinent copula densities. In particular, this approach has revealed that some standard bivariate models, e.g. with heavy-tailed marginals, have the first Fourier coefficient to be 0 or very close to 0. This has exhibited that no linear trend is present in such models and explained why powers of standard tests of independence such as Spearman rho, e.g., break down completely in such cases.

In \cite{Cmiel} the outputs given by the measure $q$ were used to select a list of models for simulation. Recall that $q$ is formed by aggregated Fourier coefficients of the underlying copula density in the product system of projected Haar functions. The list has been to a large extent inspired by \cite{Vexler17} as well as some papers in insurance and finance. Figures 1 and 2 in \cite{Cmiel} illustrate a wide range of shapes of related $q$'s. A similar approach was already applied in \cite{Ledwina14} in the context of testing for positive quadrant dependence. In our present study, we have used the list of models considered in \cite{Cmiel}.

As in \cite{Zhang} and \cite{Lee} we set $n=128$ but, in contrast to these works, we chose the standard nominal significance level $\alpha=0.05$. We took $d(128)=63$ and $t(128)=0.95$ in our statistic $T_n$. For all tests, finite-sample critical values were approximated via 100 000 MC runs. Empirical powers are based on 10 000  MC repetitions.

For an illustration of our findings, we present here three dependency scenarios among 22 considered in Appendix A.1. 
\begin{itemize}
\item [(a)] The first setting, labelled {\it Simple Regression: Step},  is specified as\\ $Y=\mathbb{1}(X \leq 0.5)+E$, where $X \sim U[0,1],\;E \sim N(0,2)$.
\item [(b)] The second, labelled {\it Heteroscedastic Regression: Linear}, is defined as \\ $Y=\sigma(X)E,\;X \sim U[1,16], \;E \sim N(0,1),\; \sigma(X) =\sqrt X$. 
\item [(c)] The third, named {\it Mixture II\;}, is the distribution of the mixture\\ (0.3)[{\it bivariate Cauchy}]+(0.7)[{\it bivariate Gaussian}], with both components related to the unit correlation matrix.
\end{itemize}

\begin{figure}[ht!]
\centering{
\includegraphics[trim = 20mm 170mm 20mm 20mm, clip, scale=0.7]{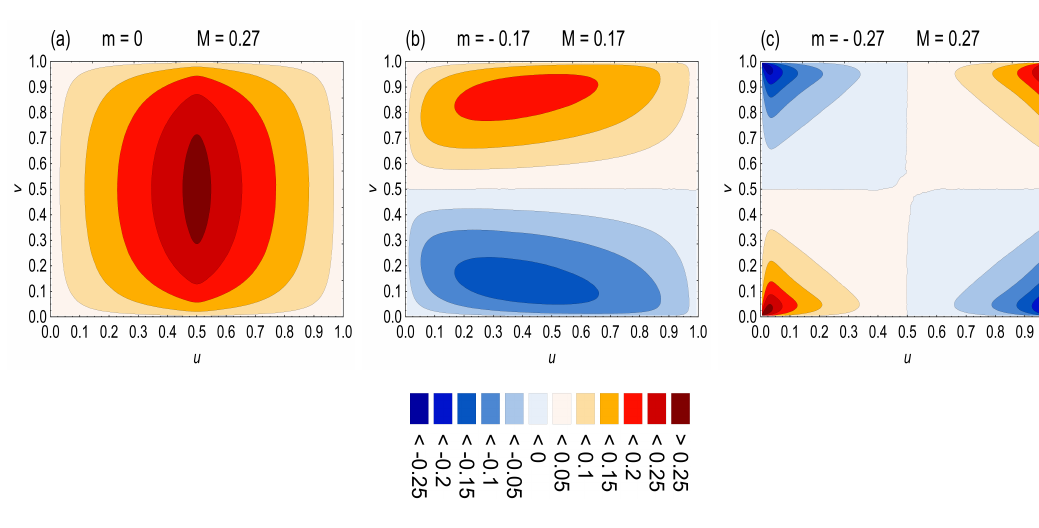}
}
\caption{Shapes of $q$ for the models (a), (b) and (c).}\label{F5}
\end{figure}

In Figure \ref{F5} we show the shapes of related $q$, estimated as an average of 100 000 empirical $\bar{q}_n$ with $n=1\ 000$ in each of the scenarios. $m$ and $M$ denote the minimal and maximal value of $q$ over $(0,1)^2$, respectively.

The above scenarios represent some typical situations among the ones considered. In addition, in selected cases, the related powers of the tests studied well reflect some general tendencies observed in the full study presented in Appendix A.1. Table 1 summarises the empirical powers (in percentages)  of the six tests in the present study. 
\begin{table}[ht!]
\caption{Powers (in $\%$) under the three dependence structures (a), (b) and (c). All tests were performed at the nominal level $5\%$.}

\begin{tabular}{ c c c c c c c c c }
 \textbf{Model} & \textbf{dCov} & \textbf{DC} &\textbf{HHG} & $\boldsymbol{\tau^*_n}$ & \textbf{Max BET}& \textbf{Ensemble BET}& $\boldsymbol{T_n}$& \textbf{HHG/}$\boldsymbol{T_n}$ \\ 
 (a) & 92& 90&86 & 92 & 66 & 68 & 87 & 87 \\  
 (b) & 47& 68&83 & 50 & 38 & 37 & 64 & 80 \\
 (c) & 7& 10&60 & 8 & 9 & 19 & 81 & 77   
\end{tabular}

\label{tab1}
\end{table}

We see that in the case of {\it Simple Regression}, where the most substantial allocation of mass is concentrated around the marginal medians,  all tests have relatively high power. However, for a more complex structure represented here by {\it Heteroscedastic Regression}, in which considerably large differences are observed near two edges,  the tests dCov, $\tau_n^*$, Max BET and Ensemble BET are evidently weaker than the competitors. Finally, when heavy tails enter via {\it Mixture II}, and the main differences between the null and actual model are manifested near the vertices of the unit square,  the tests dCov, DC, $\tau_n^*$, Max BET, and Ensemble BET are very week. 

In general, our simulation exhibits that under simple regression models, which we have considered, all tests under study behave reasonably well. However, in some relatively complex regression models, the tests dCov, DC, $\tau_n^*$, Max BET and Ensemble BET are not stable. These solutions sometimes work nicely, but in other cases lead to low powers. The situation is even more dramatic for them under several classical bivariate heavy-tailed models. For the 22 alternatives considered, we notice that the Ensemble BET test improves the construction of Max BET, while, on average, dCov and $\tau_n^*$ have slightly higher power than Ensemble
BET. DC is the best in this group of tests. 

In turn, in the whole study, the tests HHG and $T_n$ yield stable and high
powers. We have discovered a situation where $T_n$ is relatively weak (with the present
sample size). HHG is slightly more powerful than
$T_n$ in several regression models, while $T_n$ slightly outperforms HHG under classical
bivariate copulas. The statistic HHG/$T_n$ nicely exploits and matches the advantages
of both components and is the best (in terms of empirical power) test in our study. In general, our simulations exhibit that this group of tests is qualitatively different from the more classical competitors, which we have investigated, and favour HHG, $T_n$ and their combination HHG/${T_n}$ over them. 

The test statistic $T_n$ is simple and intuitive, while the
pertinent framework and tools allow us to see which data features lead to rejection of
the independence hypothesis. The simulated power in Table \ref{tab1} and Table A.1 was obtained with a smoothing parameter $t(n)$ of 0.95. Increasing or decreasing the level of smoothing in $T_n$ has a predictable effect on the associated empirical powers and allows the construction of tests that place greater emphasis on detecting very smooth or strong local deviations from independence; some evidence can be found in Appendix A.2. Moreover, it should be admitted that data processing
(e.g. estimating $p$-values) with the use of HHG and hence HHG/$T_n$ is very time consuming.
In contrast, the related computations for $T_n$ itself are very fast. For more details on the last point, see Appendix A.1. In conclusion, the above makes $T_n$ a very competitive solution.

\subsection{Application to COVID-19 infection dynamics data}\label{s5.3}

\cite{CC} have studied and compared the dynamics of COVID-19 infection between 80 countries, based on publicly available data; see \cite{Hasell}. In particular, for each country, they estimated three parameters that describe the dynamics of the first wave of infections: the percentage of confirmed infections, the time of the peak, and the duration of the first wave. They have also studied pairwise dependencies between the above parameters and the gross domestic product (GDP) per capita and population density in each country. Due to some missing data in the database, 73 out of 80 countries were used for the analysis in \cite{CC}. To visualise dependencies and study their significance, they have applied the estimator $Q^*_{n,4}$ of $q$ and the test statistic ${\cal L}_{\epsilon,6,n}^*$, $\epsilon=0.01$, proposed in \cite{Cmiel}. For a discussion of the data under consideration, some of their limitations, and related literature, we refer to \cite{CC}. 
Data used in this analysis will be made available upon request.  Figure 4 in \cite{CC} shows that in these examples the underlying dependence structure is more complex than in the two well-known examples that we have discussed in Section \ref{s4}. 

Our goal in this section is to see how our new estimate $\bar q_n$ and new test $T_n$ work for such types of data and to compare them with the other procedures considered in this paper.

\begin{figure}[ht]
\centering{
\includegraphics[trim = 17mm 130mm 0mm 0mm, clip, scale=0.7]{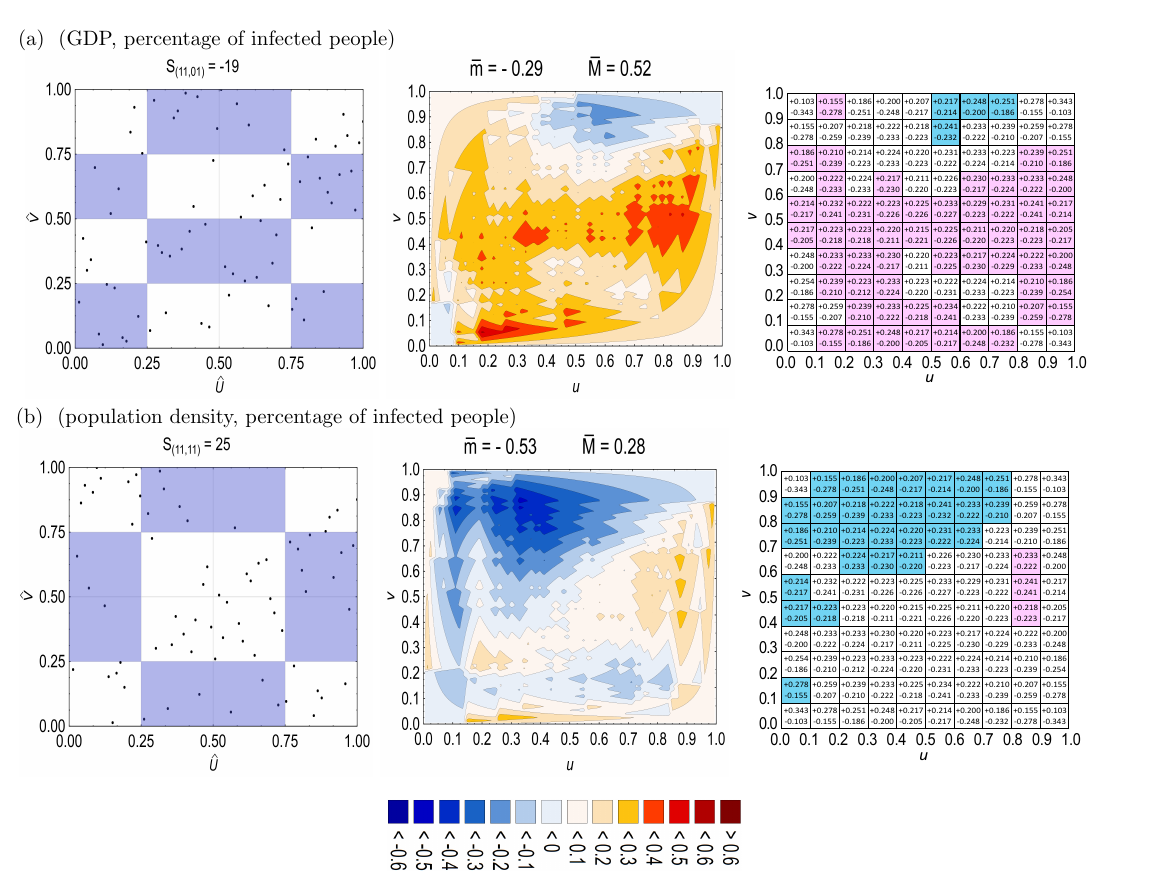}
}
\caption{Scatter plot, Zhang's pattern, estimated $q$, and the dependence diagram: (a) GDP and percentage of infected people data; (b) population density and percentage of infected people data. Barriers in dependence diagrams are based on 100 000 MC runs, $d(73)=31$.}\label{FCC}
\end{figure}

For illustration, we have selected two sets of 73 pairs of observations. These two examples are summarised in Figure \ref{FCC}. We have applied $d(73)=31$ in our discretization.

Panel (a) summarises the joint behaviour of the observations (GDP, percentage of infected people).
It seems that the plot of $\bar{q}_n$ does not suggest a simple general pattern of the allocation of the probability mass. However, several clusters of significantly large local departures from
independence are exhibited by it. These local dependencies are strong enough to reject the
global independence hypothesis at the 0.05 level by all tests except Max BET and Ensemble BET.
Zhang's approach indicates the sum of blue rectangles in panel (a) as the underlying
dependence structure. The related $p$-value is equal to $0.279$.

For observations of (population density, percentage of infected people), displayed in
panel (b) via the scatter plot, empirical $q$ reveals a considerable allocation of mass from
the upper left corner of the unit square. This allocation exemplifies the standard understanding of the
concept of negative association in the respective region. Any of nine patterns proposed in \cite{Zhang} do not reflect
well such a strongly asymmetric structure, as this one revealed by $\bar{q}_n$ and related dependence
diagram. Anyway, selecting the Max BET symmetric pattern, represented in panel (b) by the sum of white squares, leads to a relatively small $p$-value equal to 0.043. In case (b), all tests, except dCov, $\tau_n^*$ and Max BET,  invalidate global independence at the standard level $\alpha = 0.05$.

Table \ref{tabCC} collects the $p$-values for the global independence hypothesis pertinent to all the tests considered in this section. We have applied $T_n$ with $d(n)=31$ and $t(n)=0.95$.
\newpage
\begin{table}[ht!]
\caption{$p$-values for COVID-19 infection dynamics data. Based on 100 000 MC runs. $p$-values smaller than $\alpha=0.05$ are printed in read colour.}

\begin{tabular}{ c c c c c c c c c }
 \textbf{Data} & \textbf{dCov}& \textbf{DC}& \textbf{HHG} & $\boldsymbol{\tau^*_n}$ & \textbf{Max BET}& \textbf{Ensemble BET}& $\boldsymbol{T_n}$& \textbf{HHG/}$\boldsymbol{T_n}$ \\ 
 (a) &\textcolor{red}{0.008} &\textcolor{red}{0.014} & \textcolor{red}{0.005} & \textcolor{red}{0.008} & 0.250 & 0.250 & \textcolor{red}{0.035} & \textcolor{red}{0.010}   \\
 (b) &0.068 &\textcolor{red}{0.035} & \textcolor{red}{0.011} & 0.092 & 0.154 & \textcolor{red}{0.010} & \textcolor{red}{0.004} & \textcolor{red}{0.006}  
\end{tabular}

\label{tabCC}
\end{table}

For more examples of analysis of real-word datasets see \cite{Cmiel24}. 

\section{Multivariate variants of $q$ and $\bar{q}_n$}\label{s6}
Consider a random vector $\mathbf{X}=(X^{(1)},...,X^{(m)})$ with continuous univariate margins. Let $\mathbf{X}_1,...,\mathbf{X}_n$ be an i.i.d. sample, where each $\mathbf{X}_i=(X_i^{(1)},...,X_i^{(m)})$ has the same distribution as $\mathbf{X}$. Let $C$ be the unique copula pertinent to the joint cdf of $\mathbf{X}$. Let $\mathbf{X}^{(k)}=(X_1^{(k)},...,X_n^{(k)})$, $k=1,...,m$. Denote by $R_i^{(k)}$ the rank of $X_i^{(k)}$ among $X_1^{(k)},...,X_n^{(k)}$ and $\mathbf{R}^{(k)}=(R_1^{(k)},...,R_n^{(k)})$. 

The standard estimator of $C$ has the form  
$$
C_n(t_1,...,t_m)=\frac{1}{n}\sum\limits_{i=1}^n \prod\limits_{k=1}^m \mathbb{1}\left(\frac{R^{(k)}_i}{n} \leq t_k \right), \ \ \ \ (t_1,...,t_m)\in[0,1]^m,
$$
see \cite{Deheuvels_80}. We consider the following continuous and multilinear version of $\bar{C}_n$ defined in Section \ref{s3}
$$
\bar{C}_n(t_1,...,t_m)= \sum\limits_{(j_1,...,j_m)\in \{0,1\}^m} C_n\left(\frac{\lfloor nt_1 \rfloor + j_1}{n},...,\frac{\lfloor nt_m \rfloor + j_m}{n}\right) \prod\limits_{k=1}^m|nt_k+j_k-\lfloor nt_k \rfloor -1| .
$$
An equivalent formula, labelled by $\nu_n$, can be found in \cite{Deheuvels_80}.  
By elementary calculations
$$
\bar{C}_n(t_1,...,t_m)=\frac{1}{n}\sum_{i=1}^n  \sum\limits_{(j_1,...,j_m)\in \{0,1\}^m} \prod\limits_{k=1}^m \Bigl\{ \mathbb{1}(R^{(k)}_i  \leq \lfloor nt_k \rfloor + j_k)\ |nt_k+j_k-\lfloor nt_k \rfloor -1|\Bigr\}
$$
$$
=\frac{1}{n}\sum_{i=1}^n \prod\limits_{k=1}^m \Bigl\{\mathbb{1}(R^{(k)}_i \leq  \lfloor nt_k \rfloor)(1-nt_k + \lfloor nt_k \rfloor) + \mathbb{1}(R^{(k)}_i \leq \lfloor nt_k \rfloor +1) (nt_k- \lfloor nt_k \rfloor)\Bigr\}
$$
$$
=\frac{1}{n}\sum_{i=1}^n \prod\limits_{k=1}^m \Bigl\{\mathbb{1}(R^{(k)}_i \leq  \lfloor nt_k \rfloor) + (nt_k- \lfloor nt_k \rfloor)\mathbb{1}(R^{(k)}_i=\lfloor nt_k \rfloor +1)\Bigr\}.
$$
This is a counterpart of (\ref{6}) in Section \ref{s3.1}. One can also check that $\bar{C}_n$ is unbiased under independence, i.e.
$$\mathrm{E}_{{\mathbb H}_0} \bar{C}_n(t_1,...,t_m) = \prod\limits_{k=1}^m t_k .$$ 
Under fixed $n$, the related expression for the variance of $\bar{C}_n$ is quite complicated. Fortunately, by Theorem 3 form \cite{Deheuvels_80}, we have 
$$\lim\limits_{n\rightarrow \infty} \frac{\mathrm{Var}_{{\mathbb H}_0} C_n(t_1,...,t_m)}{\sqrt{n^{-1} \sigma^2(t_1,...,t_m)}}=\lim\limits_{n\rightarrow \infty} \frac{\mathrm{Var}_{{\mathbb H}_0} \bar{C}_n(t_1,...,t_m)}{\sqrt{n^{-1} \sigma^2(t_1,...,t_m)}}=1, 
$$
where 
$$\sigma^2(t_1,...,t_m)=\left(\prod\limits_{k=1}^m t_k\right) \left(1+(m-1)\prod\limits_{k=1}^m t_k- \sum\limits_{k=1}^m \prod\limits_{j\neq k} t_j\right).$$
Moreover, under ${\mathbb H}_0$ and fixed $(t_1,...,t_m)$, $(t_1,...,t_m)\in(0,1]^m$, $\sum\limits_{k=1}^m \mathbb{1}(t_k=1)<m-1$, it holds that
$$\frac{\bar{C}_n(t_1,...,t_m)-\prod\limits_{k=1}^m t_k}{\sqrt{n^{-1} \sigma^2(t_1,...,t_m)}}\xrightarrow[n\rightarrow \infty]{\mathcal{D}} N(0,1),$$
where $\xrightarrow[]{\mathcal{D}}$ denotes convergence in distribution. 

In view of the above, we define the multivariate version of the quantile dependence function as 
$$q(t_1,...,t_m)=\frac{C(t_1,...,t_m)-\prod\limits_{k=1}^m t_k}{\sigma(t_1,...,t_m)},$$
while its continuous multilinear estimator is set to be
$$\bar{q}_n(t_1,...,t_m)=\frac{\bar{C}_n(t_1,...,t_m)-\prod\limits_{k=1}^m t_k}{\sigma(t_1,...,t_m)}.$$

\section{Conclusion and discussion}\label{s7}

We have proposed a novel copula-based framework for studying the dependence structure at different quantile levels of two continuous outcomes. The proposed approach features direct interpretation and visualisation and easy implementation. One of the new tools are simultaneous local acceptance regions which provide a statistically meaningful evaluation of the strength of local departures from independence. Another is an intuitive and powerful test of the global independence hypothesis. All procedures are based on ranks, which makes them robust to outliers, invariant under strictly increasing transformations of the data, and computationally efficient.

Some extensions of the proposed approach would be of interest. We mention a few of them. The framework and related solutions can be easily adjusted to one-sided testing problem on existence of a PQD structure, the problem being important in insurance, for example. The results of \cite{Ledwina14} are encouraging in this respect. Also, studying some extensions of our test in the multivariate setting could be useful. One possibility is to use and further elaborate an application of multivariate extension of $\bar q_n$ in a similar way as done in \cite{Genest19} for more traditional statistical functionals. Another possibility is to use recent developments in statistical applications of optimal transport theory, as exemplified by \cite{Shi25}. Finally, note that our test $T_n$ is naturally suited to multivariate independence testing via combinations of pairwise rank correlations. The approach has been intensively studied in recent years, especially in high-dimensional situations, as it is well suited to enhance the versatility and power across different correlation structures, covering both sparse and dense alternative models; see, for example, \cite{Han25} for recent developments and extensive discussion.

\



{\small
\bibliographystyle{elsarticle-harv} 
\bibliography{paper-ref.bib}%
}



\normalsize

\topmargin=-2cm
\pagestyle{headings}
\textwidth 6.5in
\textheight 25cm
\hoffset -2cm

\newpage
\begin{spacing}{1.2}
\setcounter{page}{1}
\begin{center}
{\Large {\tt Supplementary Material}}\\
\vspace{1cm}
\end{center}

\noindent
{\sf A.1. Full list of alternatives, empirical powers, and related comments}\\

\noindent
We use the list of models proposed in \'Cmiel $\&$ Ledwina (2020). It has been defined as follows.\\
\\
\noindent
{\bf Simple Regression}:
\begin{itemize}
\vspace{-0.4cm}
\item[SR1:] {\it Linear}\ \ \  
$\;\;Y=2+X+\epsilon,\ \;X\sim U[0,1],\ \;\epsilon \sim N(0,1)$;
\vspace{-0.4cm}
\item[SR2:] {\it Root} \ \ 
$\;\;Y=X^{1/4} +\epsilon,\ \;X\sim U[0,1],\ \;\epsilon \sim N(0,0.25)$;
\vspace{-0.4cm}
\item[SR3:] {\it Step} \ 
$\;\;Y=\mathbb{1}(X\leq0.5) + \epsilon,\;X\sim U[0,1],\ \epsilon \sim N(0,2)$;  
\vspace{-0.4cm}
\item[SR4:] {\it Logarithmic} 
$\;\;Y=\log(1+|X|) + \epsilon,\;X \sim N(0,1);\epsilon \sim N(0,1)$;  
\vspace{-0.4cm}
\item[SR5:] {\it W-shaped} 
$\;\;Y=4[(2X-1)^2 -0.5]^2+\epsilon,\;X\sim U[0,1],\;\epsilon \sim N(0,0.5)$;  
\end{itemize}
\noindent
\hspace{-0.08cm}\textbf{Heterosceadestic Regression}:
\begin{itemize}
\vspace{-0.4cm}
\item[HR1:] {\it  Reciprocal} 
$\;\;Y=\sigma(X)\epsilon,\;$ $X$ has an exponential distribution with $\lambda=0.1$, $\epsilon \sim N(0,1),\;\sigma(X)=\sqrt{1+ 1/ X^2}$;
\vspace{-0.4cm}
\item[HR2:] {\it Linear  } 
$\;\;Y=\sigma(X)\epsilon,\;$ $X \sim U[1,16],\;$ $\epsilon \sim N(0,1),\;\sigma(X)=\sqrt X$.
\end{itemize}
\noindent
\hspace{-0.05cm}{\bf Random-Effect-Type Models}:
\begin{itemize}
\vspace{-0.4cm}
\item[RE1:] {\it Linear}\  
$\;\;Y=2+X+\epsilon_M X + \epsilon_A,\ \;X\sim U[0,1],\ \;\epsilon_M \sim N(0,4),\ \;\epsilon_A \sim N(0,1)$;
\vspace{-0.4cm}
\item[RE2:] {\it Quadratic}  
$\;\;Y=\epsilon_M(2+X+X^2)+\epsilon_A,\;\epsilon_M \sim N(0,1),\;\epsilon_A \sim N(0,1)$;
\vspace{-0.4cm}
\item[RE3:] {\it Reciprocal} 
$\;\;Y=\epsilon_M X^{-1}+\epsilon_A,\;\epsilon_M \sim N(0,1),\;\epsilon_A \sim N(0,1)$;
\vspace{-0.4cm}
\item[RE4:] {\it Heavy tailed}  
$\;(X_0,Y_0)$ is a bivariate Cauchy. Define $X=X_0, Y=\epsilon_M Y_0 +\epsilon_A$, $\epsilon_M \sim N(0,1),\;\epsilon_A \sim N(0,1)$. 
\end{itemize}
\hspace{-0.12cm}{\bf  Classical Bivariate Models}: 
\begin{itemize}
\vspace{-0.4cm}
\item[BM1:] {\it Gaussian}\;\; 
bivariate normal distribution with  $\rho=0.3$;
\vspace{-0.4cm}
\item[BM2:] {\it Mixture I}\;\; 
the mixture (0.1)(standard bivariate Gaussian) + (0.9)(bivariate Gaussian with mean 0, variances 6 and covariance 5);
\vspace{-0.4cm}
\item[BM3:] {\it Mixture II}\;\; 
the mixture (0.3)(bivariate Cauchy) + (0.7)(standard bivariate Gaussian), pertaining to the unit correlation matrix;
\vspace{-0.4cm}
\item[BM4:] {\it Switched regression}\;\; 
$Y=\mu(X) + \epsilon,\;\epsilon \sim N(0,1), X\sim N(0,1),\;\mu(X)=0$ for $|X|\leq 1.96$ and $\mu(X)=-X$ otherwise; 
\vspace{-0.4cm}
\item[BM5:] {\it Mai-Scherer copula}$\;\;$ 
Mai-Scherer distribution with parameters $(0.92,0.08)$;  
\vspace{-0.4cm}
\item[BM6:] {\it Gumbel copula}$\;\;$ 
Gumbel bivariate exponential distribution with $\theta=0.5$;
\vspace{-0.4cm}
\item[BM7:] {\it Gumbel-Hougaard copula}\;\; 
Gumbel-Hougaard distribution with $\theta=1.2$;
\vspace{-0.4cm}
\item[BM8:] {\it Cauchy}\;\; 
bivariate Cauchy distribution, pertaining to the unit correlation matrix; 
\vspace{-0.4cm}
\item[BM9:] {\it Student symmetric}\;\; 
symmetric Student's distribution with 2 degrees of freedom, pertaining to the unit correlation matrix;
\vspace{-0.4cm}
\item[BM10:] {\it Student skew}\;\; 
skew bivariate Student's distribution with 5 degrees of freedom and parameters $(0.3,0.7,-0.7)$;
\vspace{-0.4cm}
\item[BM11:] {\it Sub-Gaussian}\;\;
bivariate sub-Gaussian distribution with parameters $(0.1,1.5)$.
\end{itemize}

More details on the alternatives and the graphical presentation of $q$ for these bivariate distributions can be found in \'Cmiel $\&$ Ledwina (2020). The cases SR3, HR2, and BM3 have already been considered in Section 5.2 of the present paper. Alternatives SR1 and BM9 are graphically displayed in Appendix A.2.

We compare dCov (Sz\'ekely $\&$ Rizzo, 2009), DC (Reddi $\&$ P\'oczos, 2013) the HHG test (Heller et al., 2013), sign covariance $\tau_n^*$ (Bergsma $\&$ Dassioss, 2014), Max BET with $d_{max}=4$ (Zhang, 2019), and its improvement - the Ensemble BET test with $d_{max}=4$ (Lee et al., 2023), and our solution $T_n$ with $d(n)=63$ and $t(n)=0.95$. We have also considered the minimum $p$-value of the combination of HHG and $T_n$, denoted by HHG$/T_n$. We took $n=128$, $\alpha=0.05$. Empirical powers are based on 10 000 MC runs. Table A.1 presents the results obtained. 
\\
\\
\noindent
{\bf Table A.1:} Empirical powers under the full list of models. Based on 10 000 MC runs.
\begin{center}
\begin{figure}[h!]
\hspace*{0.1cm}
\includegraphics[trim = 10mm 110mm 10mm 20mm, clip, scale=0.87]{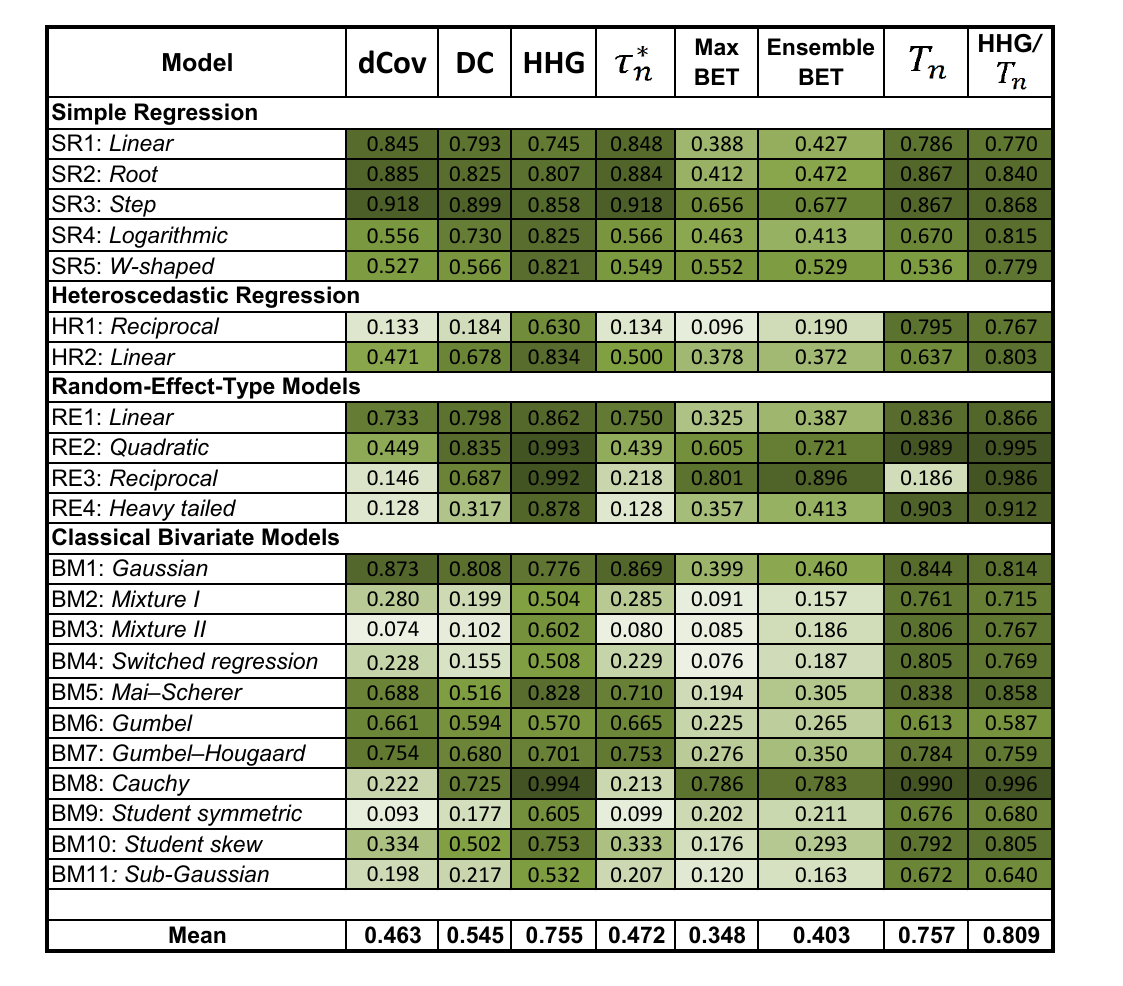}

\end{figure}
\end{center}
\vspace*{-1cm}
\noindent
Table A.1 appears to provide sufficient evidence to evaluate the tests under study. The test statistics dCov, DC, $\tau_n^*$, Max BET and Ensemble BET exhibit unstable behaviour. Though work nicely in some situations, in some others we notice a breakdown of empirical powers. In addition, we can observe striking similarities in the empirical powers of dCov and $\tau_n^*$. DC is an evident leader in this group of tests. 

The remaining statistics are qualitatively different. HHG is very stable and has high power, as a rule. It is seen to be relatively weak for BM4, only.
$T_n$ is also stable and powerful, except the alternative  RE3. 
The combination HHG/$T_n$ eliminated all weak sides of its components.

The proposed procedure $T_n$ has computational complexity $O(n^2 \log n)$. We empirically
investigate the growth of the computation time, starting from a sample size of
$n = 100$ and successively doubling the sample size. Increasing the sample size from
$n = 100$ to $n = 200$ results in an approximately 2.2-fold increase in the computation
time; from $n = 200$ to $n = 400$, a 2.3-fold increase; from $n = 400$ to $n = 800$, a
2.9-fold increase; from $n = 800$ to $n = 1 600$, a 3.7-fold increase; and from $n = 1 600$
to $n = 3 200$, a 3.9-fold increase. The asymptotic scaling factor for a doubled sample
size is 4. For $n = 100$, the computation of the test statistic with MC = 10 000 Monte
Carlo replications requires only a few seconds; specifically, on a machine equipped
with an Intel(R) Core i7-6700K CPU running at 4.00 GHz, the computation takes
approximately 6 seconds.

Upon a request of one of the referees, we have included in the revised version empirical powers of the distance covariance and one, selected by us,  kernel test.  Also, the above discussion on computational issues and the next Sub-section A.2 have resulted as a replay to the referee requirements. \\

\noindent
{\sf A.2. Some comments on the sensitivity of empirical powers to the grid size $d(n)$ and the smoothing parameter $t(n)$ }\\

We start with a simple illustration of the empirical behaviour of $V_n$ under $\mathbb{H}_0$, for $n=128$ and three choices of $d(n): 15, 31, 63$, as it clearly  exhibits some sources of problems with finite sample behaviour of weighted sup-type statistic related to the process $|\bar Q_n(u,v)|$. Recall that 
$$
\bar Q_n(u,v)=\sqrt n \bar q_n(u,v)= \sqrt n \frac{\bar C_n(u,v)-uv}{\sqrt{uv(1-u)(1-v)}},
$$
while
$$
V_n=V_n(d(n))=\max_{1 \leq j,j' \leq d(n)} \big|\bar Q_n(p_{s(n),j},p_{s(n),j'})\big|.
$$
\newpage

\vspace*{-1.5cm}
\begin{center}
\begin{figure}[h!]
\hspace*{-1.5cm}
\includegraphics[trim = 5mm 15mm 0mm 25mm, clip, scale=0.92]{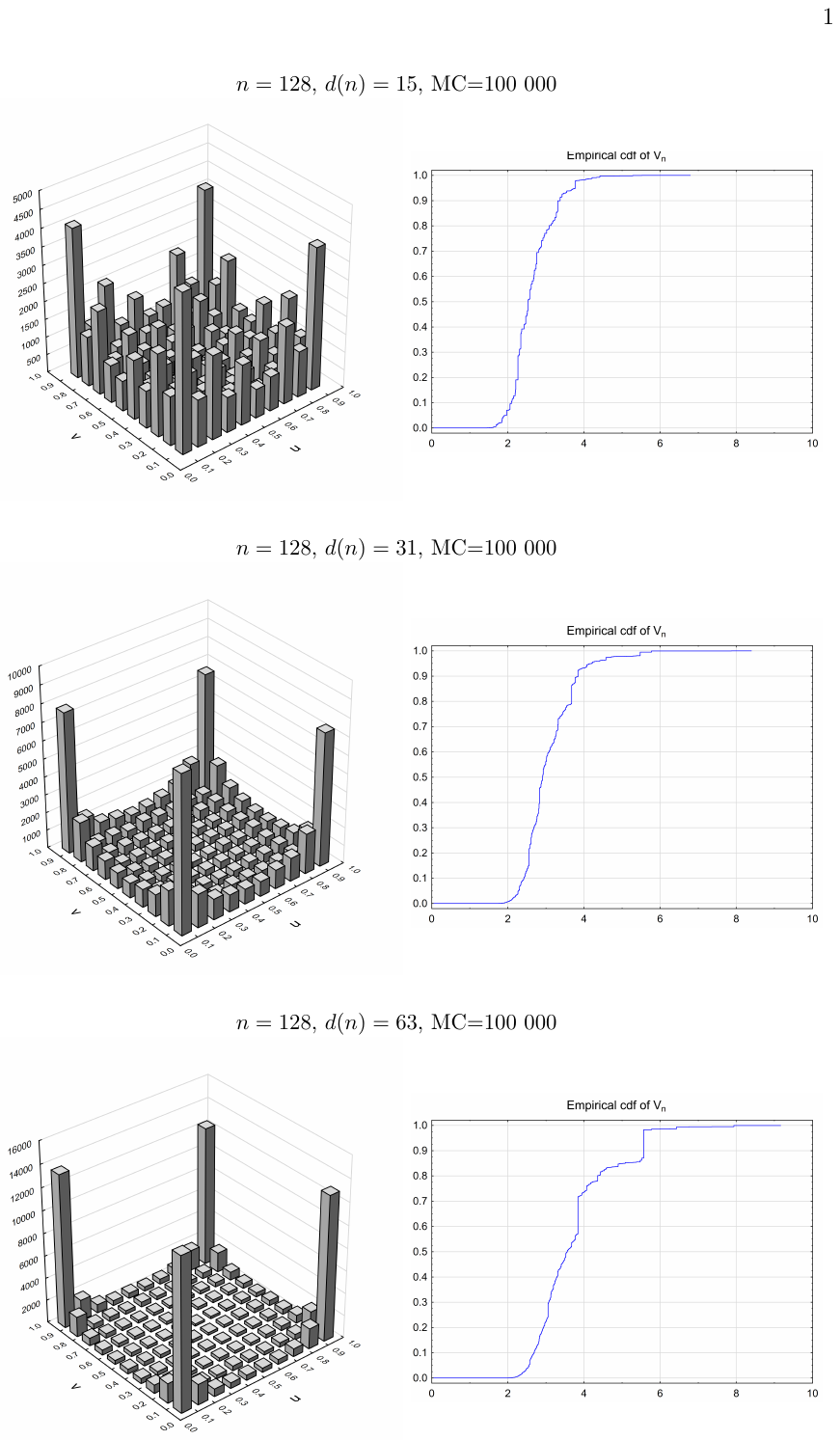}

\end{figure}
\end{center}
\vspace*{-1cm}
\noindent
{\bf Fig. A.1.} Left column: empirical distribution (histogram) of the argument that maximises $|\bar{Q}|$ on the grid against $d(n)$. Right column: empirical distribution (empirical cdf) of $V_n=V_n(d(n))$ against $d(n)$. The simulation in both columns was performed under $\mathbb{H}_0$.\\
\\

In the left column of Figure A.1 we show three histograms of the values of the arguments $(p_{s(n),j},p_{s(n),j'}),\;1 \leq j,j' \leq d(n)$, which, in the case where $\mathbb{H}_0$ is true, maximise $|\bar Q_n|$ over the selected grid. In the right column of this figure we display related empirical cdf's of $V_n$. Two observations are immediate. The rate at which these histograms  tend to one concentrated in four vertexes of the grid, as $d(n)$ is growing,  is high and accompanied with evidently increasing non-smoothness  of the empirical cdf's of related $V_n$'s and increasing heaviness of their tails. Both tendencies are unfavourable to pertinent empirical critical values, and hence to empirical powers, as well. Moreover, their simultaneousness and intensity make the problem difficult to handle. 

Our idea is as follows. Instead of investigating the maximum of $|\bar Q_n(u,v)|$, in the grid, which evidently overestimates the unknown maximum of $|q|$, we shall estimate this maximum using the statistic
$$
T_n=T(d(n),t(n))=\frac{1}{d^2(n) - \lceil t(n)d^2(n)\rceil +1} \sum_{k =\lceil t(n)d^2(n)\rceil} ^{ d^2(n)} \big|\bar Q\big|_{(k)}
$$ 
in which the truncation point is explicitly related to the magnitude of $d(n)$ through the number $\kappa(n)= \lceil t(n)d^2(n)\rceil$. Recall also that $\big|\bar Q \big|_{(1)} \leq ... \leq \big|\bar Q\big|_{(d^2(n))}$ are ordered values of $|\bar Q_n(u,v)|$ on the points on the grid.

The asymptotic analysis of the influence of the pair $(d(n),t(n))$ on the behaviour of sizes and powers of the test based on $T_n$ is beyond the scope of the present contribution. However, we provide some numerical evidence that the proposed test characterises stable and predictable finite sample behaviour under different choices of $(d(n),t(n))$.

We start our discussion with the case of $n=128$, already introduced in Section 5 of the paper and further studied in the previous Appendix. In Figure A.2 we show empirical critical values, pertaining to levels $\alpha=0.05$, $\alpha=0.01$, and two choices of $t(n)$: $t(n)=0.95$, $t(n)=0.99$.
\begin{center}
\begin{figure}[h!]
\vspace*{-1.0cm}
\includegraphics[trim = 0mm 170mm 0mm 25mm, clip, scale=0.79]{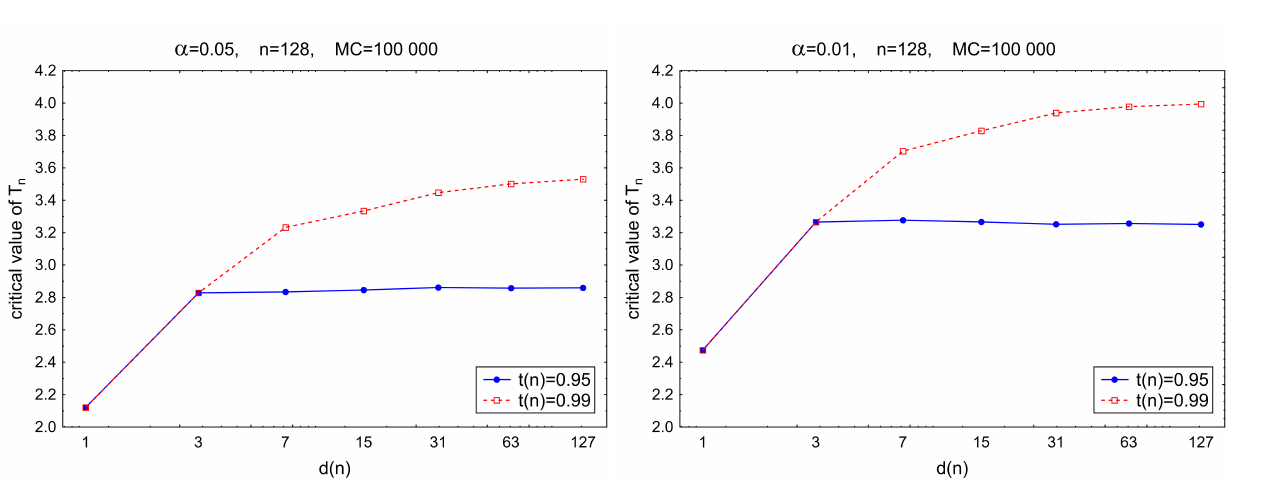}

\end{figure}
\end{center}
\vspace*{-1.5cm}
\noindent
{\bf Fig. A.2.} Critical values of $T_n=T_n(d(n),t(n))$ for levels $\alpha=0.05$ (left column) and $\alpha=0.01$ (right column) versus $d(n)$ (logarithmic scale) and $t(n)$.
\newpage
We see that empirical critical values behave stably starting from a reasonably small $d(n)=7$ and changes in their magnitude, as $d(n)$ is growing, are very moderate.
Moreover, a high degree of smoothing ($t(n)=0.95$) makes the pertinent critical values practically insensitive to changes of $d(n)$.

Now, we shall illustrate how, given the parameters $n=128$ and $d(n)=63$, used in the simulation of powers, the empirical powers vary depending on the three choices of $t(n): 0.90, 0.95, 0.99$. To simplify the presentation, we have numbered 1,...,22 successive alternatives appearing in rows of Table A.1, staring from the top to the bottom.  
\begin{center}
\begin{figure}[h!]
\vspace*{-1.0cm}
\includegraphics[trim = 3mm 170mm 0mm 25mm, clip, scale=0.89]{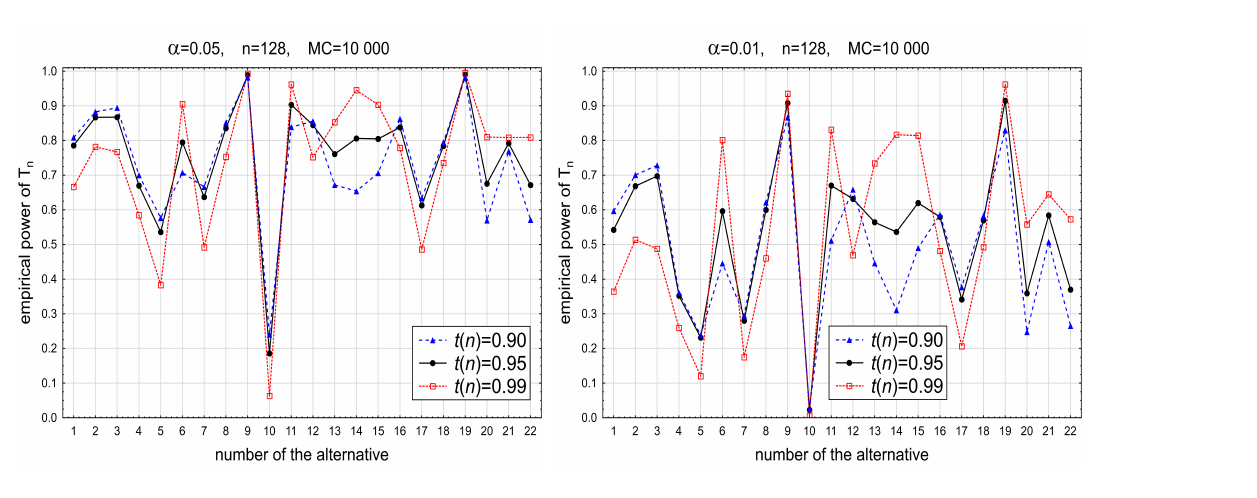}

\end{figure}
\end{center}
\vspace*{-1.5cm}
\noindent
{\bf Fig. A.3.} Empirical powers of $T_n=T_n(d(n),t(n))$ at levels $\alpha=0.05$ (left column) and $\alpha=0.01$ (right column), under 22 alternatives from Table A.1, against $t(n)$.\\
\\
We see that the empirical powers pertaining to $t(n)=0.90$ and $t(n)=0.99$ differ considerably in many cases, while the choice $t(n)=0.95$ yields intermediate results. To better understand the reaction of $T_n=T_n(d(n),t(n))$ to the choice of $t(n)$ let us consider two alternatives 1 (SR1) and 14 (BM9), representing the global and local allocation of mass. The pertinent shapes of $q$ are shown in Figure A.4. 

\begin{figure}[h!]
\vspace*{-0.7cm}
\center{
\includegraphics[trim = 3mm 160mm 0mm 25mm, clip, scale=0.70]{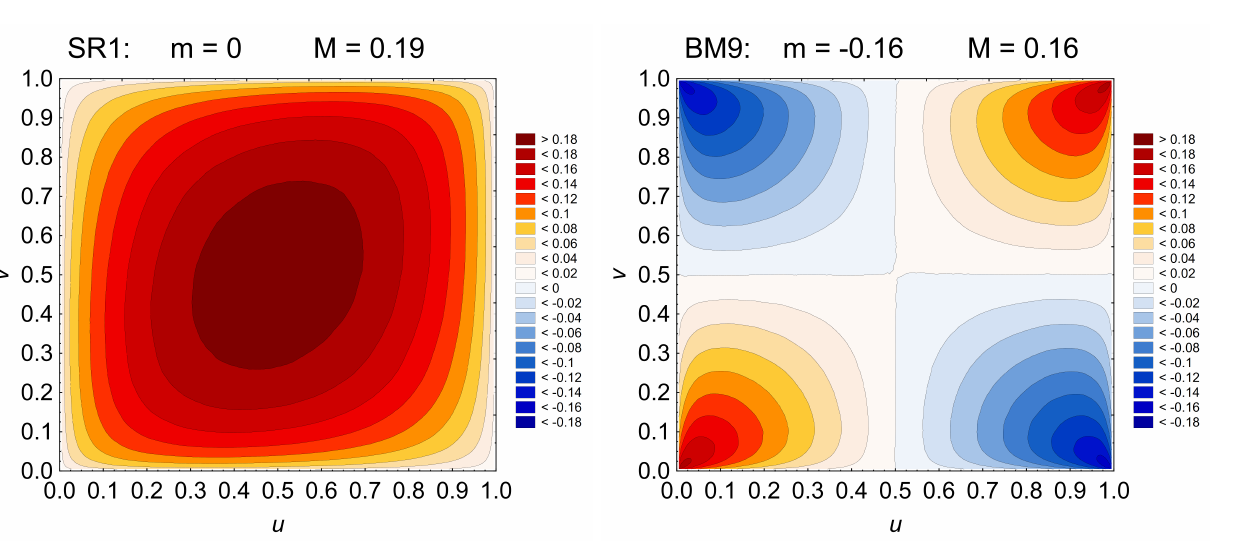}
}
\end{figure}

\vspace*{-0.7cm}
\noindent
{\bf Fig. A.4.} Shapes of $q$ in the case of the alternative SR1 (simple linear regression) and BM9 (Student symmetric). $m$ and $M$ denote the minimum and maximum value of $q$, respectively.
\newpage
Looking at $q$ for the first alternative (simple linear regression), it is intuitive that it is possible and  profitable to sum up  many (potentially) available relatively large estimated values of such $q$ (i.e. use $t(n)=0.90$), while in the case of alternative 14 (symmetric Student), it is much better to include to $T_n$ only essentially large estimated values of $q$ (i.e. use $t(n)=0.99$), despite the fact that pertaining critical value is larger than it would be under $t(n)=0.95$ and $t(n)=0.90$.
A similar interpretation applies to other alternatives. The pertinent shapes of $q$ can be found in \'Cmiel and Ledwina (2020).

The above discussion shows that $T_n(t(n),d(n))$ is a flexible solution that can be easily adjusted to some particular needs (e.g., achieving high sensitivity to local deviations from independence) through an adequate choice of $d(n)$ and $t(n)$. In the following, we propose some choices of these parameters that aim to balance the sensitivity of the associated test to local and global departures from independence.

Based on our experience, the suggested formulas for $d(n)$ and $t(n)$ are as follows
$$d(n)=2^{\lfloor \log_2(n/2) \rfloor}-1,\ \ \  t(n)=1-1/(2\sqrt{n}).$$
Table A.2 contains critical values of $T_n(d(n),t(n))$ for two standard significance levels, six exemplary sample sizes, related $t(n)$'s, calculated with use of the above formula, and several values of $d(n)$'s in each case. For each $n$ and both significance levels, critical values obtained with use of the pair $(t(n),d(n))$ defined above are printed in bold face.\\
\\
\noindent
{\bf Table A.2:} Critical values of $T_n=T_n(d(n),t(n))$ for the levels $\alpha=0.05$ and $\alpha=0.01$ against $d(n)$.
\begin{table}[ht!]
\center{
\begin{tabular}{ c|c c c c c c c c c }
\hline
          &  \multicolumn{9}{|c}{ $d(n)$}      \\ 
\hline
                &$1$      & $3$     & $7$   & $15$   & $31$  & $63$   & $127$   & $255$ & $511$ \\
\ \ $n/t(n)/\alpha$  &  \multicolumn{9}{|c}{ critical values of $T_n$}      \\ 
\hline
$100/0.95/0.05$ & $2.00$  & $2.54$ & $2.80$ & $2.82$& $\mathbf{2.82}$ & $2.83$ &         &        &  \\
$100/0.95/0.01$ & $2.40$  & $3.07$ & $3.23$ & $3.22$& $\mathbf{3.22}$ & $3.22$ &         &        &  \\
\hline
$200/0.96/0.05$ & $1.98$  & $2.61$ & $2.95$ & $2.94$& $2.93$ & $\mathbf{2.92}$ & $2.92$  &        &  \\
$200/0.96/0.01$ & $2.53$  & $3.21$ & $3.38$ & $3.36$& $3.33$ & $\mathbf{3.33}$ & $3.33$  &        &  \\
\hline
$300/0.97/0.05$ & $1.98$  & $2.69$ & $2.96$ & $3.02$& $3.02$ & $3.03$ & $\mathbf{3.03}$  & $3.04$ &  \\
$300/0.97/0.01$ & $2.54$  & $3.23$ & $3.41$ & $3.43$& $3.44$ & $3.44$ & $\mathbf{3.44}$  & $3.46$ &  \\
\hline
$400/0.98/0.05$ & $1.98$  & $2.67$ & $3.13$ & $3.13$& $3.14$ & $3.15$ & $\mathbf{3.15}$  & $3.15$ &  \\
$400/0.98/0.01$ & $2.57$  & $3.23$ & $3.56$ & $3.56$& $3.57$ & $3.57$ & $\mathbf{3.57}$  & $3.58$ &  \\
\hline
$500/0.98/0.05$ & $1.96$  & $2.68$ & $3.14$ & $3.13$& $3.14$ & $3.15$ & $\mathbf{3.15}$  & $3.15$ &  \\
$500/0.98/0.01$ & $2.58$  & $3.20$ & $3.56$ & $3.56$& $3.56$ & $3.56$ & $\mathbf{3.57}$  & $3.58$ &  \\
\hline
$600/0.98/0.05$ & $1.96$  & $2.70$ & $3.14$ & $3.14$& $3.15$ & $3.15$ & $3.15$  & $\mathbf{3.15}$ & $3.16$  \\
$600/0.98/0.01$ & $2.58$  & $3.20$ & $3.56$ & $3.55$& $3.56$ & $3.56$ & $3.56$  & $\mathbf{3.56}$ & $3.58$ \\
\end{tabular}
}
\end{table}

In Table A.2 one can see the same phenomenon with respect to the critical values of $T_n$ as
exhibited in Figure A.2. Namely, given $n$, for $d(n) \geq 7$, the careful choice of the parameter $t(n)$ practically
stops related critical values of $T_n(d(n),t(n))$ from increasing with $d(n)$.
\\

\noindent
{\sf A.3. The values of barriers in selected real data examples and their stability} \\

In this section, we reproduce on a larger scale the barriers that appear in our two real data examples. The sample sizes are: 88, 517 while $d(n)$ are: 31, 255, accordingly.  Throughout, we consider $\Pi_{k,l},\; k,l=1,...,10$, as defined in Section 4.1. The barriers $\ell^{-}(n,\alpha,\Pi_{k,l})$ and 
$\ell^{+}(n,\alpha,\Pi_{k,l})$ are calculated for $\alpha=0.05$ and are shown in Fig. A.5.

\vspace*{-0.5cm}
\begin{center}
\begin{figure}[h!]
\hspace*{-1.5cm}
\includegraphics[trim = 10mm 180mm 0mm 20mm, clip, scale=1]{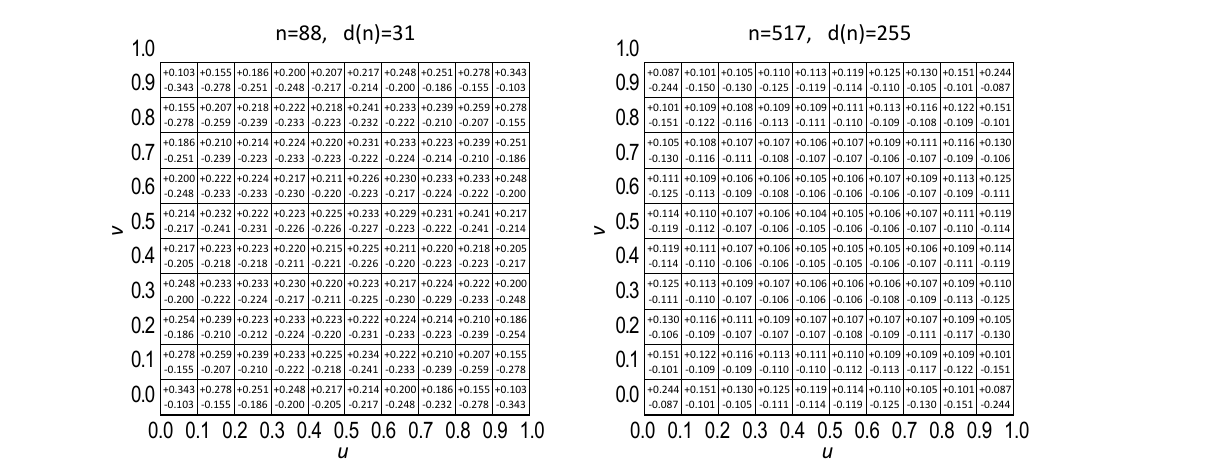}

\end{figure}
\end{center}
\vspace*{-1cm}
\noindent
{\bf Fig. A.5.} Barriers in local acceptance regions against sample size $n$ and $d(n)$. Based on 100 000 MC runs.\\
\\
We start with simple observation concerning some symmetries seen in Fig. A.5. Since $\bar q_n$ obeys the properties of the quantile dependence function, therefore, in particular, it satisfies properties 8 - 10 of Proposition 1 in \'Cmiel and Ledwina (2020). The barriers are calculated using the realisations of $(U,V),\;U \sim U[0,1], V \sim U[0,1]$. However, on the other hand, the pairs or random variables  $(U,V),\; (U,1-V),\; (1-U,V)$ and $(1-U,1-V)$,  have the same distribution under the independence of $U$ and $V$.  Hence, due to the mentioned properties of $\bar q_n$, several symmetries and asymmetries can be noticed in these tables.

Now, let us briefly comment on magnitudes of these empirical barriers. Recall that each barrier pertains to a single cell $\Pi_{k,l},\;k,l=1,...,10,$ and a related subset of points of the grid, $P_{k,l}$, which fall into the cell $\Pi_{k,l}$. Therefore, unpleasant effects related to the suprema of the weighted empirical process, which we have illustrated in Figure A.1, appear here in a much smaller scale. In real-world examples, which we have considered in the paper, the situation with stability of barriers is easier than the one considered in the first row of Figure A.1. The common feature of both situations is that the problem is more difficult for cells located in the vertices and along the edges of $[0,1]^2$. This circumstance influences values of barriers in these regions. However, even for a relatively small sample size $n=88$, we found the results satisfactory. What is much more important, as $n$ is increasing,  fast growth of precision, especially near the edges,  is noticed. Namely, we see that the barriers 
$\ell^{-}(n,\alpha,\Pi_{k,l})$ and $\ell^{+}(n,\alpha,\Pi_{k,l})$ are getting noticeably closer as $n$ increases, in spite of the fact that $d(n)$ increases as well.

In turn, in Fig. A.6, given $n$,  we illustrate in more detail the influence of growing $d(n)$ on the resulting barriers. We restrict our attention to $n=88$ and $n=517$, three levels of $d(n)$: 31, 67, and 127, and elements lying on the main diagonal of the respective matrix of results. Again, it is seen that as $n$ is growing, considerable stability of the results is achieved. In any case, even with a relatively small sample size $n=88$, the choice $d(n)=31$ yields reasonably stable results.   
\begin{figure}[h!]
\vspace{-0.7cm}
\centering{
\includegraphics[trim = 20mm 60mm 0mm 20mm, clip, scale=0.98]{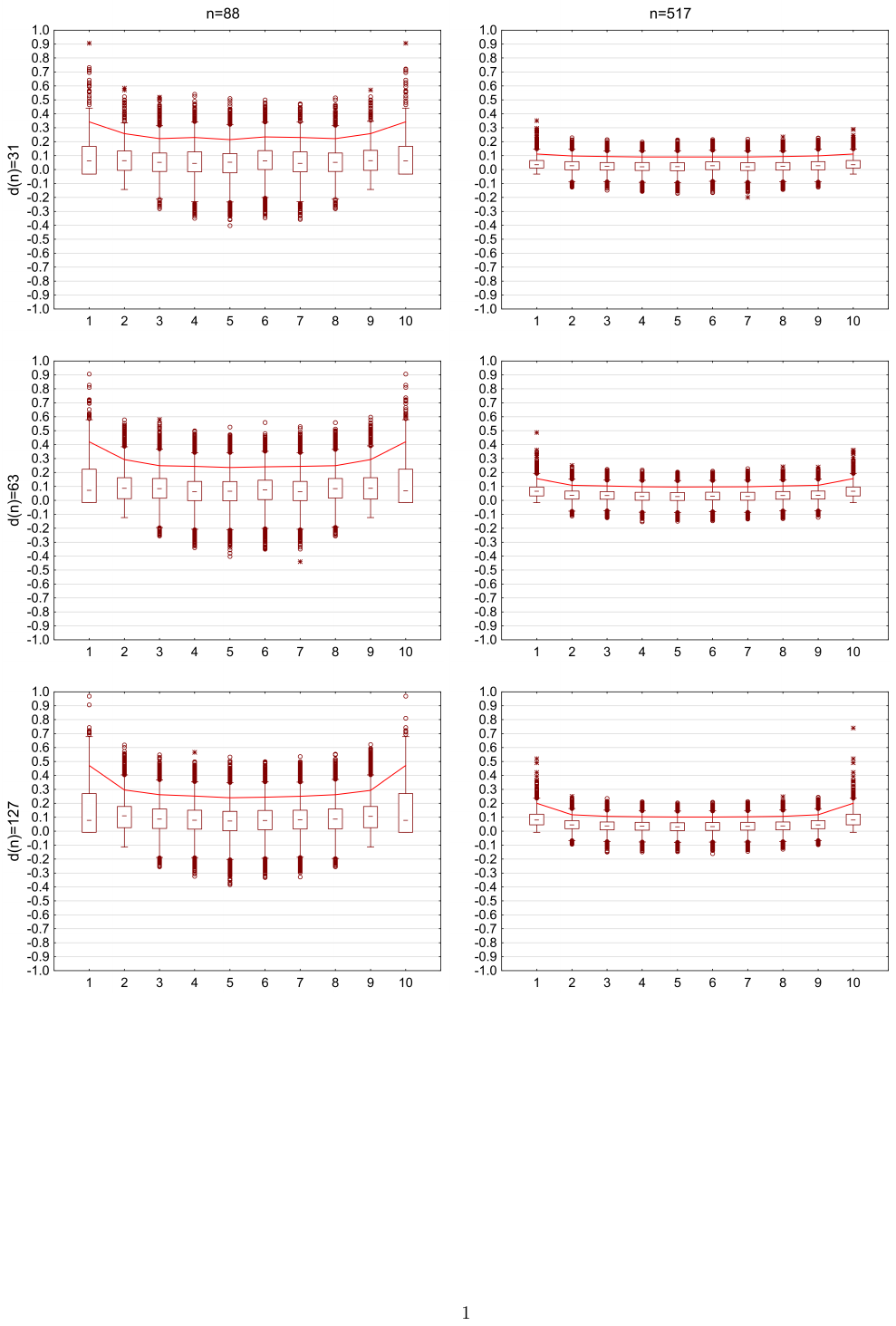}
}
\end{figure} \vspace{-2.0cm}
\\
\noindent 
{\bf Fig. A.6.} Bar plots for $\ell^{+}(n,\alpha,\Pi_{k,k}),\;k=1,...,10,$ for $n=88$ and $n=517$ against $d(n)=31, 63, 127$. Based on 100 000 MC repetitions.\\

As mentioned in Section 4, if one wants to consider very dense grids, then the construction of barriers should be improved. 
\\
\\
\noindent
{\sf A.4. Proofs}\\

\noindent
{\it A.4.1. Proof of Proposition 1}\\
By least squares formula for linear regression
$$\frac{z_2-\mathrm{E}(\mathbb{1}(Y>y_v))}{\sqrt{\mathrm{Var}(\mathbb{1}(Y>y_v))}}=\mathrm{Corr}(\mathbb{1}(Y>y_v),\mathbb{1}(X>x_u))\frac{z_1-\mathrm{E}(\mathbb{1}(X>x_u))}{\sqrt{\mathrm{Var}(\mathbb{1}(X>x_u))}}.$$ 
But  
$$\mathrm{Corr}(\mathbb{1}(Y>y_v),\mathbb{1}(X>x_u))=\frac{P(Y>y_v,X>x_u)-(1-u)(1-v)}{\sqrt{u(1-u)v(1-v)}}=q(u,v).$$
Moreover,
$$\mathrm{E}(\mathbb{1}(Y>y_v)|\mathbb{1}(X>x_u)=0)=P(Y>y_v|X<x_u),$$
$$\mathrm{E}(\mathbb{1}(Y>y_v)|\mathbb{1}(X>x_u)=1)=P(Y>y_v|X>x_u),$$
so the slope of the line passing through the points $(0,E(\mathbb{1}(Y>y_v)|\mathbb{1}(X>x_u)=0)$ and $(1,E(\mathbb{1}(Y>y_v)|\mathbb{1}(X>x_u)=1)$ is equal to $P(Y>y_v|X>x_u)-P(Y>y_v|X\leq x_u)$. Since this line is in fact the linear regression, the second part of Proposition 1 is proven. The third part follows by properties of bivariate cdf and elementary transformations. 
 \hfill{$\Box$}
\\
\\
\noindent
{\it A.4.2. Proof of Proposition 2}\\
In formula (3) we have four terms, each of the structure $k_j(w_j,z_j)C_n(w_j,z_j)$,  with adequately defined arguments $(w_j,z_j)$ and non-negative coefficients $k_j(w_j,z_j) , \;j=1,...,4$. The sample size $n$ is fixed and, to increase readability,  is neglected in the above notation.

First, consider the classical estimate $C_n(u,v)$. We have
$$
C_n(u,v)=\frac{1}{n} \sum_{i=1}^n \mathbb{1}\left(\frac{R_i}{n} \leq u, \frac{S_i}{n} \leq v\right)=
\frac{1}{n} \sum_{i=1}^n \mathbb{1}\left(\frac{R_i}{n} \leq \frac{\lfloor nu \rfloor}{n}, \frac{S_i}{n} \leq \frac{\lfloor nv \rfloor}{n}\right) =
$$
$$
\frac{1}{n} \sum_{i=1}^n \mathbb{1}\left(\frac{R_i}{n} \leq w_1, \frac{S_i}{n} \leq z_1\right)=w_1+z_1-1+\frac{1}{n}\sum_{i=1}^n \mathbb{1}\left(\frac{R_i}{n} \geq w_4, \frac{S_i}{n} \geq z_4\right)=
$$
$$
w_1+z_1-1+\frac{1}{n}\sum_{i=1}^n \mathbb{1}_{[w_4,1] \times [z_4,1]}\left(\frac{R_i}{n}, \frac{S_i}{n}\right)=
w_1+z_1-1+ \int_{{\mathbb R}^2} \mathbb{1}_{[w_4,1] \times [z_4,1]} (F_n(x),G_n(y))dH_n(x,y).
$$

Given $w_4$ and $z_4$, the function $\mathbb{1}_{[w_4,1] \times [z_4,1]}(u,v)$ is right-continuous and lattice-superadditive; cf. R\"{u}schendorf (1980), e.g. Recall also that we restrict attention to bivariate cdf's $H$ with continuous marginals $F$ and $G$. In such circumstances, Theorem 4.1 of Schriever (1987), along with the comment following it, implies that $C_n$ is stochastically ordered in the sense of Proposition 2. In this way, we also have an alternative simple proof of Proposition 1 in Ledwina and Wy{\l}upek (2014).

In view of the structure of $\bar C_n(u,v)$ given by $\sum_{j=1}^4 k_j(w_j,z_j)C_n(w_j,z_j)$, a similar argument applies to $\bar C_n(u,v)$ and thus concludes the proof of Proposition 2. \hfill{$\Box$}
\\
\\
\noindent
{\it A.4.3. Proof of Theorem 1}\\
Recall that $d(s)=2^{s+1}-1,$ and $p_{s,1}=1/(1+d(s))$. For a short set $c(n)=p_{s(n),1}$. Due to our assumptions, $c(n) \to 0$ and $\sqrt n c(n) \to \infty$. Moreover, we have
$$
T_n \leq  \sup\limits_{(u,v)\in[c(n),1-c(n)]^2}\ \ \sqrt n|\bar{q}_n(u,v)|. 
$$
On the other hand, 
$$\bar{q}_n(u,v)=\frac{\bar{C}_n(u,v)-uv}{\sqrt{uv(1-u)(1-v)}}$$
$$=\frac{\bar{C}_n(u,v)-C_n(u,v)}{\sqrt{uv(1-u)(1-v)}}+\frac{C_n(u,v)-C(u,v)}{\sqrt{uv(1-u)(1-v)}}+\frac{C(u,v)-uv}{\sqrt{uv(1-u)(1-v)}}$$
$$:=A_1(u,v)+A_2(u,v)+A_3(u,v).$$
Hence
$$
T_n \leq  \sup\limits_{(u,v)\in[c(n),1-c(n)]^2}\sqrt n |A_1(u,v)|+\sup\limits_{(u,v)\in[c(n),1-c(n)]^2}\sqrt n|A_2(u,v)|+\sup\limits_{(u,v)\in[c(n),1-c(n)]^2}\sqrt n|A_3(u,v)|.
$$
Since $|\bar{C}_n(u,v)-C_n(u,v)|\leq 2/n$ with probability 1, we have
$$
\sup\limits_{(u,v)\in[c(n),1-c(n)]^2}\sqrt n |A_1(u,v)|\leq \sqrt{n}\frac{2/n}{c(n)(1-c(n))}=\frac{2}{\sqrt{n}c(n)(1-c(n))},
$$
with probability 1. The condition $\sqrt{n}c(n)\rightarrow \infty$ yields
$$
\sup\limits_{(u,v)\in[c(n),1-c(n)]^2}\sqrt n |A_1(u,v)|\rightarrow 0, \eqno(A.1)
$$
with probability 1. 

By Proposition 3.1 of Segers (2012), under our assumptions,  $\sqrt n |C_n(u,v)-C(u,v)|$ tends weakly, in the space $\ell^{\infty}([0,1])^2$, to the Gaussian process (3.1) therein. This implies that
$$\label{B.2}
\sup\limits_{(u,v)\in[c(n),1-c(n)]^2}\sqrt n |A_2(u,v)|=O_P(2^{s(n)})=O_P(d(n)). \eqno(A.2)
$$

When the null hypothesis is valid, then $A_3(u,v)=0$ and we conclude that  
$$
T_n \leq O_P(d(n)) \eqno(A.3)
$$
under independence.

Now we shall show that, under the alternatives, $T_n$ tends to infinity faster than under the null model. For this purpose recall that for a function $J$ on $(0,1)^2$ we denote by $|J|_{(k)},\; k=1,...,[d(n)]^2 $ the $k$-th order statistic among $|J(p_{s(n),j},p_{s(n),j'})|,\;j,j'=1,...,d(n)$. With these notations, write
$$
T_n=\frac{1}{K(n)-\kappa(n)+1}\sum\limits_{k=\kappa(n)}^{K(n)}|\bar{Q}|_{(k)}=\frac{1}{K(n)-\kappa(n)+1}\sum\limits_{k=\kappa(n)}^{K(n)}\sqrt{n}|A_1+A_2+A_3|_{(k)}
$$
$$
\geq \frac{1}{K(n)-\kappa(n)+1}\left(\sum\limits_{k=\kappa(n)}^{K(n)}\sqrt{n}|A_3|_{(k)}-\sum\limits_{k=\kappa(n)}^{K(n)}\sqrt{n}|A_2|_{(k)}-\sum\limits_{k=\kappa(n)}^{K(n)}\sqrt{n}|A_1|_{(k)}\right)
$$
$$
\hspace{-7.5cm}\geq \frac{1}{K(n)-\kappa(n)+1}\sum\limits_{k=\kappa(n)}^{K(n)}\sqrt{n}|A_3|_{(k)}
$$
$$
\hspace{-2.5cm}- \sup\limits_{(u,v)\in[c(n),1-c(n)]^2}\sqrt n |A_2(u,v)|\;- \sup\limits_{(u,v)\in[c(n),1-c(n)]^2}\sqrt n |A_1(u,v)|. \eqno(A.4)
$$
In view of (A.1) and (A.2), to get the result, it is enough to consider the first component in (A.4). Let $C$ represents the alternative cdf under consideration. Since $C$ is continuous on $[0,1]^2$,\;pertinent $A_3$ satisfies $A_3 \not\equiv 0$, under $C$, while $d(n) \to \infty$, as $n \to \infty$, therefore, given the alternative $C$, there exist $n_0$, $a_0>0$ and $p_0>0$, such that for all $n>n_0$ for the least $p_0 K(n)$ points in the net we have $|C(u,v)-uv|>a_0>0$. Hence,
$$
\frac{1}{K(n)-\kappa(n)+1}\sum\limits_{k=\kappa(n)}^{K(n)}\sqrt{n}|A_3|_{(k)}\geq \sqrt{n}\ \frac{a_0 \min\{p_0 K(n),K(n)-\kappa(n)+1\}}{K(n)-\kappa(n)+1}=O(\sqrt{n})
$$
This yields 
$$
T_n \geq O_P(\sqrt n) \eqno(A.5)
$$
under the alternative $C$. Since $d(n)/\sqrt n \to 0$, by the assumption, (A.3) along with (A.5) imply the consistency under $C$. \hfill{$\Box$} 
\\
\\

\noindent
{\sf A.5. Zhang's BET framework. New interpretation and related comments}\\ 

To describe the nine patterns introduced by Zhang (2019) and exploited by Xiang et al. (2023) as well as Lee et al. (2023) we shall use the Walsh (1923) system of functions. We follow the description in Paley (1932). To define the needed elements, only those that we shall use in this section, introduce the first two Rademacher functions $r_1(s)$ and $r_2(s)$, where
$$
r_j(s)=\sum_{k=1}^{2^j} (-1)^{k-1}\mathbb{1}_{[\frac{k-1}{2^j},\frac{k}{2^j})}(s),\;\;\;s \in [0,1),\;\;\;j=1,2.
$$
Next, recall the definition of the first four Walsh functions $w_0(s),\; w_1(s),\; w_2(s)$ and $w_3(s)$, $\;s \in [0,1)$. We have $\;w_0(s) \equiv 1$, while $\; w_1(s)=r_1(s), \; w_2(s)=r_2(s)$ and $w_3(s)=r_1(s)r_2(s)$.  The Walsh functions $\;w_j(s),\;j \geq 1,\;$ take on only two values +1 and -1, on respective subintervals of [0,1) induced by the Rademacher system.

Now, observe that the nine successive blue and white patterns in Fig. 2 of Xiang et al. (2023), seen as the $3 \times 3$ matrix, are represented by the nine first functions of the product system $z_{i,j}(s,t)=w_i(s)w_j(t),\;i,j =1,2,3,\;(s,t) \in [0,1)^2$. More precisely, the regions where the respective $z_{i,j}$'s are equal to -1 are marked in blue, while the white regions correspond to the value +1.

In turn, the symmetry statistics pertinent to these patterns of association are simply standard empirical Fourier coefficients of a copula density in this product system. To be specific,  denote here the symmetry statistics by $W_{i,j}$ and observe that
$$
W_{i,j}=\int_0^1 \int_0^1 w_i(s)w_j(t) d C_n(s,t),\;\;\;i,j=1,2,3, 
\eqno(A.6)
$$ 
where $C_n(s,t)$ is the classical estimate of the underlying copula $C$, as defined in Section 3. Since Walsh functions are orthonormal, (A.6) shows that $W_{i,j}$'s can also be seen as empirical correlation coefficients of relevant functions of observations. 

For readers accustomed to Zhang's notation, we link our notation to the original ones: $W_{1,1}=S_{(10,10)},\  W_{1,2}=S_{(10,01)},  \  W_{1,3}=S_{(10,11)},$
$\ W_{2,1}=S_{(01,10)},\  W_{2,2}=S_{(01,01)},  \  W_{2,3}=S_{(01,11)},$
$\ W_{3,1}=S_{(11,10)},\ W_{3,2}=S_{(11,01)},  \ W_{3,3}=S_{(11,11)}$. 

A comparison of (A.6) and (6) shows that Zhang's and ours approaches are not as different as they appear at first glance.

The selection of a pattern via Zhang's Max BET proceeds as follows. We calculate 
$$
W=\text{Max BET}=\max_{1 \leq i,j \leq 3} |W_{i,j}|
$$
and indicate the index $(i_0,j_0)$, say, for which the maximum is attained. If $W_{(i_0,j_0)} > 0$, then the sum of the pertinent white rectangles represents the selected pattern. Otherwise, the sum of related blue rectangles is the chosen template. The procedure is accompanied by the appropriate $p$-value of the selected $W_{(i_0,j_0)}$. Recall that Zhang has introduced the symmetry statistics starting from the contingency table generated by the binary expansion up to the prescribed depths. 

Observe that using the empirical Fourier coefficients $W_{i,j}$  and $W_{j,j'}^o$, given in (6) and (A.6), respectively,  we can consider the two approaches discussed above on the same ground. Namely, we can think in terms of modelling alternatives via exponential models or series expansions for a copula density and work on pertinent score vectors, which, in such a setting, are formed just by related empirical Fourier coefficients. For an illustration and discussion on an approach of such type, see Section 2 of Kallenberg $\&$ Ledwina (1999). The approach has roots in the paper by Neyman (1937), who proposed, in nonparametric goodness-of-fit problem, to introduce a list of parametric exponential models to approximate the unknown nonparametric density and developed related locally optimal solution. Neyman's paper exploited Legendre's system to build a class of auxiliary models. Hamdan (1964) applied the Welsh system to elaborate the respective variant of Neyman's smooth test for uniformity.

Viewing $W_{i,j}, W_{i,j}^o$ and other components as to empirical Fourier coefficients has an additional advantage. That is, it links independence testing to checking relations \\ $\text {Cov}\bigl(f_1(F(X)),f_2(G(Y))\bigr)=0$ for $f_1$ and $f_2$ ranging over a separating class of functions; cf. Breiman (1968), p. 165 ff. For example, the approach of Kallenberg and Ledwina (1999) has pertained to the system of Legendre's polynomials. New needs stimulated search for new classes of separating functions with better abilities for localisations of departures from independence, easier interpretation, more parsimonious description, etc. In particular, thinking in terms of Fourier coefficients allows one to see the discretizations introduced by Ledwina and Wy{\l}upek (2014), Ma and Mao (2019), and Zhang (2019) from such a perspective.
  
In our opinion, introducing the Welsh system greatly simplifies the description of Zhang's framework to fit the pattern to the data at hand. In addition, some ready results related to the system can be exploited. Needless to say that the description via the Welsh functions extends to higher depths as well. In particular, the test statistic for the global independence hypothesis proposed in Section 4.5 of Zhang (2019) and the Ensemble BET test in Lee et al. (2023), which we examine in our simulation study, can also be expressed with their use.

We close with some observations on how fitting of one of the eight Zhang patterns works in practice. On the one hand, there is extensive evidence (Zhang 2019, Xiang et al. 2023, Lee et al. 2023) that patterns are handy and useful in detecting and displaying some tendencies in allocation of pseudo-observations in $[0,1]^2$. On the other hand, it is also clear that in some situations the small set of templates is not rich enough and sufficiently flexible. Zhang (2019), p. 1632, discusses some limitation of the approach in case of local dependency. Many other situations can be easily indicated. Our examples of Section 5.3 can serve as an illustration. \\
\\
\\  
\noindent
{\sf A.6. Brief comments on some approaches to measuring association}\\ 

The literature on measuring and quantifying associations between two variables $X$ and $Y$, say, is immense. In our brief summary below, we only highlight some aspects of detecting associations. 

There are many numeric measures of dependence; see Geenens and de Micheaux (2022) for recent impact, comprehensive discussion, and related literature. Several sets of axioms imposed on such numeric measures are strongly influenced by the work of R\'enyi (1959). Although mathematically elegant, many of these axioms are simply unrealistic. For example, the assumed symmetry in $X$ and $Y$ immediately removes from consideration most regression models in which the uncertainty of one variable conditional on the second is the core of the relationship. Reimherr and Nicolae (2013) propose to considerably reduce the postulates and emphasise the need to have a clear interpretation associated with the measures of dependence. Our objection concerning numeric measures is that they are aimed at something that is simply impossible to achieve: to express a complex dependence structure via a single number.

For better description and understanding of the underlying structure of dependence, several local approaches have been introduced, and some numeric local measures have been proposed. These measures are functions of arguments $x$ and $y$ from $\mathbb{R}$. Tj{\o}stheim et al. (2022) presents comprehensively some history and the most recent developments. This stream of papers is technically involved, as a rule.

Still another stream of investigations concerns some function-valued measures of dependence, which typically are functions of some auxiliary variables, often quantile levels. Kowalczyk and Pleszczy\'nska (1977), Coles et al. (1999), and
Li et al. (2017) can serve as an illustration. Formally, also several local measures fall into this category, the construction of Bjerve and Doksum (1993) being a prominent example. An extensive discussion of local measures from this perspective can be found in Drouet Mari and Kotz (2001).

Our emphasis is on a measure with simple interpretation and that being capable of exhibiting different dependence structures. Existing evidence shows that the measure $q$ introduced in Ledwina (2014, 2015) satisfies such requirements. It is a function-valued measure of dependence, defined in $(0,1)^2$, and is explicitly related to the so-called quantile hits popular in econometrics. We believe that the tools related to this measure, which we have developed in this paper, further support its use. \vspace{3cm}
\end{spacing}

\noindent
{\tt References for Appendices}\\
\begin{spacing}{1}  
\begin{description}[topsep=0pt,itemsep=0pt,parsep=0pt,labelsep=0em]	
\item Bergsma, W. $\&$ Dassios, A. (2014). A consistent test of independence based on a sign covariance related to Kendall's tau. {\it Bernoulli} {\bf  20}, 1006-1028.
\item Bjerve, S. $\&$   Doksum, K. (1993). Correlation curves: Measures of association as functions of covariate values. {\it Annals of Statistics} {\bf  21}, 890-902.
\item Breiman, L. (1968). {\it Probability}. Reading, MA: Addison-Wesley.
\item \'Cmiel, B. $\&$ Ledwina, T. (2020). Validation of association. {\it Insurance: Mathematics and Economics} {\bf 91}, 55-67.
\item Coles, S. $\&$ Heffernan, J. $\&$ Tawn, J. (1999). Dependence measures for extreme value analyses. {\it Extremes} {\bf 2}, 339-365.
\item Drouet Mari, D. $\&$ Kotz, S. (2001). Correlation and Dependence. Imperial College Press. London.
\item Geenens, G. $\&$ de Micheaux, P. L. (2022). The Hellinger correlation. {\it Journal of the American Statistical Association} {\bf 117}, 639-653.
\item Hamdan, M.A. (1964). A smooth test of goodness of fit based on the Walsh functions. {\it Australian Journal of Statistics} {\bf 6}, 130-136.
\item Heller, R., Heller, Y. $\&$ Gorfine, M. (2013). A consistent multivariate test of association based on ranks of differences. {\it Biometrika} {\bf 100}, 503-510.
\item {Kallenberg, W.C.M. $\&$ Ledwina, T.} {(1999)}. {{Data-driven rank tests for independence}}, {\it Journal of the American Statistical Association} {\bf 94}, {285-301}.
\item {Kowalczyk, T. $\&$ Pleszczy\'nska, E.} {(1997)}. {Monotonic dependence functions of bivariate distributions}. {\it Annals of Statistics} {\bf 5}, {1221-122}.
\item {Ledwina, T.} {(2014)}. {Dependence function for bivariate cdf's}. {\it arXiv:1405.2200v1 [stat.ME]}.
\item {Ledwina, T.} {(2015)}. {Visualizing association structure in bivariate copulas using new dependence function}. {\it Stochastic Models, Statistics and Their Applications}. Springer Proceedings in Mathematics $\&$ Statistics {\bf 122}, 19-27.
\item Ledwina, T. $\&$ Wy{\l}upek, G. (2014). Validation of positive quadrant dependence. {\it Insurance: Mathematics and Economics} {\bf 56}, 38-47.
\item Lee, D., Zhang, K. $\&$ Kosorok, M.R. (2023). The binary expansion randomized ensemble test. {\it Statistica Sinica} {\bf 33}, 2381-2403.
\item Li, R. $\&$  Cheng, Y. $\&$  Chen, Q. $\&$  Fine, J. (2017). Quantile association for bivariate survival data. {\it Biometrics} {\bf 73}, 506-516.
\item Ma, L. $\&$ Mao, J., (2019). Fisher exact scanning for dependency. {\it Journal of the American Statistical
Association} {\bf 114}, 245–258.
\item Neyman, J. (1937). 'Smooth' test for goodness of fit. {\it Skandinavisk Aktuarietidiskrift} {\bf 20}, 149-199.
\item Paley, R.E.A.C. (1932). A remarkable series of orthogonal functions (I). {\it Proceedings of the London Mathematical Society} {\bf 2}, 241-264.
\item Reddi, S. J. $\&$ P\'oczos, B. (2013). Scale invariant conditional dependence measures. {\it In International Conference on Machine Learning}. PMLR, 1355-1363.
\item Reimherr, M. $\&$ Nicolae, D. L. (2013). On quantifying dependence: A framework for developing interpretable measures. {\it Statistical Science}. {\bf 28}, 116-130.
\item R\'enyi, A. (1959). On measures of dependence. {\it Acta Mathematica Academiae Scientiarum Hungarica} {\bf 10}, 441-451.
\item R\"{u}schendorf, L. (1980). Inequalities for the expectation of $\Delta$-monotone functions. {\it Zeitschrift f\"{u}r Warscheinlichkeitstheorie und verwandte Gebiete} {\bf 54}, 341-349.
\item Schriever, B. (1987). An ordering for positive dependence. {\it Annals of Statistics} {\bf 15}, 1208-1214.
\item Segers, J. (2012). Asymptotics of empirical copula process under non-restrictive smoothness assumptions. {\it Bernoulli} {\bf 18}, 764-782.
\item Sz\'ekely, G. $\&$ Rizzo, M. L. (2009), Brownian distance covariance. {\it Annals of Applied Statistics} {\bf 3}, 1236-1265.
\item Tj{\o}stheim, D. $\&$ Otneim, H. $\&$ St{\o}ve, B. (2022). Statistical dependence: Beyond Pearson's $\rho$. {\it Statistical Science} {\bf 37}, 90-109.
\item Walsh, J.L. (1923). A closed set of normal orthogonal functions. {\it American Journal of Mathematics } {\bf 45},  5-24.
\item Xiang, S., Zhang, W., Liu, S., Hoadley, K.A., Perou, C.M.,  Zhang, K. $\&$  Marron, J.S. (2023). Pairwise nonlinear dependence
analysis of genomic data. {\it Annals of Applied Statistics} {\bf 17}, 2924-2943.
\item Zhang, K. (2019). BET on independence. {\it Journal of the American Statistical Association} {\bf 114}, 1620-1637.

\end{description}
\end{spacing}

\end{document}